\documentclass[superscriptaddress,
 reprint, amsmath,amssymb, aps,bbm, pra,longbibliography
]{revtex4-1}
\usepackage{comment}
\usepackage{graphicx}
\usepackage{dcolumn} 
\usepackage{bm}
\usepackage{graphics}
\usepackage{bbold}
\usepackage{bbm}
\usepackage{amsmath,amsthm}
\usepackage{lipsum}
\usepackage{amssymb}
\usepackage{textgreek}
\usepackage{physics}
\usepackage{amsmath}
\usepackage[skins,breakable]{tcolorbox}
\usepackage{tikz}
\usetikzlibrary{quantikz2}
\newcommand{\be}{\begin{equation}}
\newcommand{\ben}{\begin{eqnarray}}
\newcommand{\een}{\end{eqnarray}}
\newcommand{\ee}{\end{equation}}

\begin{document}
\title
{Subsystem evolution in the history state formalism} 
\author{A.\ Boette}
\affiliation{Instituto de F\'{\i}sica La Plata, CONICET, and Departamento\ de F\'{\i}sica, Universidad Nacional de La Plata,
C.C. 67, La Plata (1900), Argentina}
\author{F.\ Lomoc}
\affiliation{Instituto de F\'{\i}sica La Plata, CONICET, and Departamento\ de F\'{\i}sica, Universidad Nacional de La Plata,
C.C. 67, La Plata (1900), Argentina}
\author{N. Canosa}
\affiliation{Instituto de F\'{\i}sica La Plata, CONICET, and Departamento\ de F\'{\i}sica, Universidad Nacional de La Plata,
C.C. 67, La Plata (1900), Argentina}
\author{R. Rossignoli}
\affiliation{Instituto de F\'{\i}sica La Plata, CONICET, and Departamento\ de F\'{\i}sica, Universidad Nacional de La Plata,
C.C. 67, La Plata (1900), Argentina}
\affiliation{Comisi\'on de Investigaciones Cient\'{\i}ficas (CIC), La Plata (1900), Argentina}

\begin{abstract}
We present a relational framework for characterizing the 
evolution of quantum composite systems,  based on the correlations within a history state  describing the whole evolution relative to a quantum reference clock. While total system–clock entanglement entropy consistently measures the unitary evolution of the closed global system, a different picture emerges for the reduced state of an open subsystem, whose evolution is generally non-unitary due to environmental coupling. We introduce a general formalism that quantifies open-subsystem dynamics using a subsystem–clock mutual information derived from the history state. We demonstrate that this quantity serves as a consistent quantifier of subsystem evolution with respect to a given clock basis, can be applied to arbitrary dynamics, and reduces to system–time entanglement entropy in the global limit. Furthermore, maximizing this measure over all clock bases yields a basis-independent bound on the maximum evolution achievable relative to the clock, directly linked to the difference between the subsystem's entanglement with the rest of the history state and its entanglement with the environment. Finally, we derive analytical expressions for qubit clocks, illustrate our results using spin-$1/2$ and spin-$1$ quantum walks, and present a full matrix formalism of Kraus operators for describing the most general subsystem evolution.

 \end{abstract} 
\maketitle
\vspace*{-1.4cm}

\section{Introduction}

The history state formalism, stemming from the seminal work of Page and Wootters \cite{Paw2.83}, has garnered significant interest across diverse areas of physics, including non-relativistic quantum mechanics (QM) \cite{Ma.15,Moreva.14,Clean.15,BRGC.16,Pa.17,BR.18,Ni.18,Ll.19,DP.19,Cir.22,LBCR.22,DN.25,L.26}, quantum gravity, and relativistic QM \cite{CR.91,IS.94,Ga.09,zh.18,di.19,Diaz.212,G.23,DR.26}. Within this relational framework, time is incorporated via a reference quantum clock, and system evolution emerges from an entangled system--clock history state satisfying a timeless Wheeler--DeWitt-like equation \cite{dewitt.1967}. Beyond its foundational significance, this approach enables efficient quantum simulation schemes requiring only $\log N$ qubits to encode $N$ time steps \cite{BRGC.16,DN.25}. Moreover,  history states can be generated using quantum circuits suitable for optical implementations \cite{DP.19}. 

In previous works \cite{BRGC.16,BR.18}, we showed that the system--time entanglement entropy of the history state serves as a consistent quantifier of the degree of unitary evolution of a closed system. We subsequently applied this formalism to one-dimensional (1D) spin-$1/2$ Quantum Walks (QW) \cite{LBCR.22} to quantify the effective number of orthogonal states visited along the walk, and recently extended it to time-series analysis through a quantum embedding  \cite{L.26}. 

However, when considering a composite quantum system
$S\,\text{=}\,A\,\text{+}\,B$, a fundamental question arises: how can we properly quantify the degree of evolution of a subsystem $A$ within this relational framework? When $A$ is coupled to an environment $B$ its reduced state $\rho_A(t) ={\rm Tr}_B\,[\rho_S(t)]$ is generally mixed and undergoes non-unitary dynamics, even when the global system evolves unitarily \cite{KR.83,NC.00,Zurek.03,Peb.03}. While such subsystem dynamics can be described by a standard completely positive trace-preserving (CPTP) map when the initial global state is a product state $\rho_A(0)\otimes \rho_B(0)$ \cite{KR.83,NC.00}, this characterization breaks down when initial system--environment correlations or entanglement are present \cite{Pck.94,Kimura.03,Terno.08,Modi.12,Bu.14, modi.15,Lu.16}. 

Moreover, as we show in this manuscript, the entanglement of the subsystem--clock reduced state can no longer be directly interpreted, in general, as a measure of the degree of evolution of $\rho_A(t)$. Instead, we introduce a formalism based on a subsystem--clock mutual information obtained following an unread local measurement on the clock, which consistently quantifies the extent of evolution of $\rho_A(t)$ for arbitrary initial conditions. Formally, this measure corresponds to a Holevo quantity given by the difference between the total subsystem--rest entanglement and the average subsystem--environment entanglement. It is  an indicator of the effective number of orthogonal subsystem states visited throughout the history and reduces  to the global system--time entanglement entropy when $A$  encompasses the entire system (as for a trivial or  ``frozen'' environment).  Otherwise, it is upper-bounded by the full system--time entanglement entropy, such that the evolution of $\rho_A(t)$ never ``exceeds'' that of the global state $\rho_S(t)$.

Furthermore, whereas for a closed global system such mutual information is strictly independent of the choice of the clock basis, for an open subsystem it generally depends on this choice, reflecting that the degree of reduced-state evolution depends on the reference clock basis. Maximizing this quantity over all clock bases yields a basis-independent measure of the maximum subsystem evolution achievable relative to the clock $T$, which is directly related to the difference between the total $A\text{--}BT$  entanglement and the $A\text{--}B$  entanglement alone. In this context,  the $A\text{--}T$ entanglement governs the sensitivity of the subsystem's evolution to the clock basis: this basis dependence grows as $A\text{--}T$ entanglement decreases, reaching its maximum when the $A\text{--}T$  reduced state is  classically correlated (hence with no entanglement).

The  paper is organized as follows. In Sec.~\ref{sec:formalism}, we first review the main features of the history state formalism along with system--time entanglement and mutual information, and then introduce our general framework for quantifying open-subsystem evolution, for  arbitrary initial states and clock bases. In Sec.~\ref{sec:examples}, we illustrate these results first for a qubit clock, deriving analytical expressions for the evolution quantifier and its basis dependence, and then for 1D spin-$1/2$ and spin-$1$ QW, with the aim of quantifying the evolution of the spin and position ``subsystems''.  Finally, conclusions and perspectives are provided in Sec.~\ref{sec:conclusions}, while Appendices~A and~B contain detailed derivations and proofs of results in Sec.~\ref{sec:formalism}.

\section{Formalism}\label{sec:formalism}
\subsection{History states}
We consider a closed quantum bipartite system $S=A+B$ and a reference clock system $T$,  whose states belong to  Hilbert spaces ${\cal H}_S={\cal H}_A\otimes {\cal H}_B$,  and ${\cal H}_T$ respectively. 
We assume that $S$  
is in  a pure state  $|\psi_S(t)\rangle\equiv \ket{\psi_{AB}(t)}$ 
if  $T$ is in state $|t\rangle$. 
The history state $|\Psi\rangle\in{\cal H}_S\otimes {\cal H}_T$ associated to $N$ discrete times  $t=0,\ldots,N-1$, corresponding to $N$ distinguishable  clock states $|t\rangle$, $\langle t|t'\rangle=\delta_{tt'}$, is 
\begin{equation}|\Psi\rangle=
\tfrac{1}{\sqrt{N}}\sum_{t=0}^{N-1} |\psi_S(t)\rangle \otimes |t\rangle\,.  \label{HS1}
\end{equation}
This state  describes the  evolution along these $N$ times and can be generated through a quantum circuit involving $\log N$ qubits, 
thus allowing 
an efficient evaluation of time averages and related quantities \cite{BRGC.16,BR.18,DN.25}. 

If $U(t)\equiv U_{AB}(t)$ is the unitary operator which connects $\ket{\psi_S(t)}$ to the  initial state,  
$\ket{\psi_S(t)}=U(t)\ket{\psi_S(0)}$, the state \eqref{HS1}  is an eigenstate with eigenvalue $1$  of the cyclic global 
unitary translation operator ${\cal U}$ on ${\cal H}_S\otimes {\cal H}_T$:   
\begin{align}  {\cal U}|\Psi\rangle&=|\Psi\rangle\,,\label{2}\\
{\cal U}&=\sum_{t=1}^{N} U(t)U^\dag(t-1)\otimes |t\rangle\langle t-1|,\label{3}\end{align}
 where we have set $U(N)=U(0)=\mathbb{1}$ and $|N\rangle=|0\rangle$. Hence, $|\Psi\rangle$ remains fully invariant under ${\cal U}$.  
 
 Conversely, if Eqs.\ \eqref{2}--\eqref{3} are fulfilled,   $\ket{\Psi}$ has necessarily the form \eqref{HS1} for  $\ket{\psi_S(t)}=\sqrt{N}\braket{t}{\Psi}=U(t)\ket{\psi_S(0)}$  \footnote{Here $\bra{t}(\ket{\psi}\otimes\ket{\phi}):=
 \ket{\psi}\braket{t}{\phi}\in {\cal H}_S$, hence  $\ketbra{t}(\ket{\psi}\otimes\ket{\phi})
 :=\mathbb{1}\otimes\ketbra{t}(\ket{\psi}\otimes\ket{\phi})=\braket{t}{\phi}\ket{\psi}\otimes\ket{t}$.}:  Eq.\  \eqref{2} implies $\braket{t}{\Psi}=\bra{t}{{\cal U}}\ket{\Psi}\,\forall\,t$,  i.e.\  $\ket{\psi_S(t)}=U(t)U^\dag(t\!-\!1)\ket{\psi_S(t\!-\!1)}$, which leads to $\ket{\psi_S(t)}=U(t)\ket{\psi_S(0)}$ and hence to $\ket{\Psi}=\sum_t\ketbra{t}\!\Psi\rangle=
 \frac{1}{\sqrt{N}}\!\sum_t\ket{\psi_S(t)}\otimes \ket{t}$, with $\ket{\psi_S(t)}$ normalized  if $\braket{\Psi}{\Psi}=1$. 

We also  mention that {\it any}   state $|\Psi\rangle\in {\cal H}_S\otimes {\cal H}_T$ with Schmidt rank $n_s\leq N$ can be written in the  history form \eqref{HS1} ({\it history state representation}) after a Fourier transform (or in general a complex Hadamard transform) of the clock   states of its Schmidt epresentation (App.\ A). 

\subsection{System-time entanglement and mutual information}
The entanglement of the history state \eqref{HS1}, to be denoted as system-time entanglement $E(S,T)$,  constitutes a measure of the ``amount'' of evolution of system $S$ along the $N$ times $t$, i.e., a  measure of the effective number of orthogonal states of ${\cal H}_S$ visited in this evolution. For stationary states $|\psi_S(t)\rangle=e^{-i\phi(t)}|\psi_S(0)\rangle$, $|\Psi\rangle=|\psi_S(0)\rangle(\frac{1}{\sqrt{N}}\sum_t e^{-i\phi(t)}|t\rangle)$ becomes {\it separable}, and hence $E(S,T)=0$. 
On the other hand, if a new orthogonal state is visited at each step, such that $\langle\psi_S(t)|\psi_S(t')\rangle=\delta_{tt'}$,  Eq.\ \eqref{HS1} is already its Schmidt decomposition and $|\Psi\rangle$ becomes {\it maximally entangled}, with $E(S,T)=\log N$. 

 Setting $\rho_{ST}=|\Psi\rangle\langle\Psi|$, the reduced density matrices of systems $S$ and $T$ in the history state \eqref{HS1}   are, respectively,  
\begin{equation}
\begin{split}
\rho_{S} =  {\rm Tr}_{T}\, \rho_{ST}&=\frac{1}{N}\sum_{t} |\psi_S(t)\rangle\langle\psi_S(t)|\,,\\
\hspace*{-2cm}\rho_T =  {\rm Tr}_{S}\, \rho_{ST}&=
\frac{1}{N}\sum_{t,t'}\langle \psi_S(t)|\psi_S(t')\rangle|t'\rangle\langle t| \label{e04}, 
\end{split}
\end{equation}
which are isospectral, with $\rho_S$ representing the time average of $\rho_S(t)=\ketbra{\psi_S(t)}$ along the history. The  system-time entanglement entropy can then be quantified through their common entropy, 
\begin{equation}E(S,T)=S(\rho_{S})=S(\rho_T), \label{e4}\end{equation} 
with  $S(\rho)=-{\rm Tr}\, \rho\log\rho\,$   the von Neumann entropy. This quantity then constitutes  a measure of distinguishable evolution, i.e.\ of  the effective number of distinct orthogonal states of $S$ visited in the history, which can be defined as $2^{E(S,T)}$ for $\log=\log_2$. It is fully determined by the overlap matrix of elements $\langle\psi_S(t')|\psi_S(t)\rangle$ between the system  evolved states, and takes into account the relative weight  in the history of each orthogonal state visited.  
 As in any pure state, the associated {\it mutual information} \cite{NC.00}, 
\begin{equation}
\begin{split}
I(S,T)&=
S(\rho_{ST}||\rho_S\otimes\rho_T)=S(\rho_S)-S(S|T)\\&=S(\rho_S)+S(\rho_T)-S(\rho_{ST})\,,\label{42}
\end{split}\end{equation}
where $S(S|T)=S(\rho_{ST})-S(\rho_T)$ is the {\it conditional entropy}, is just twice the entanglement entropy: 
\begin{equation}
I(S,T)=2 E(S,T)\,.\label{ESTI}
\end{equation}

Let us now consider a quantum operation based on local rank one orthogonal projectors on the standard clock basis, 
$\Pi_t=\mathbbm{1}_{S}\otimes|t\rangle\langle t|$, such that $\rho_{ST}\rightarrow \rho'_{ST}$, with 
\begin{eqnarray} \rho'_{ST}&=&{\cal E}(\rho_{ST})=\sum_t \Pi_t\rho_{ST}\, \Pi_t\nonumber\\&=&\tfrac{1}{N}\sum_t\rho_{S}(t)\otimes |t\rangle\langle t|\,,\label{S}\end{eqnarray}
 where $\rho_{S}(t)=|\psi_S(t)\rangle\langle\psi_S(t)|$. 
This is equivalent to an unread  measurement at the clock in this basis. 
The ensuing reduced states of $S$ and $T$ are $\rho'_{S}={\rm Tr}_{T}\rho'_{ST}=\frac{1}{N}\sum_t\rho_S(t)=\rho_{S}$ (no signaling)  and $\rho'_T={\rm Tr}_{S}\rho'_{ST}=\tfrac{1}{N}\mathbbm{1}_T$. The previous $S-T$ entanglement is destroyed by this operation, as the state \eqref{S} is separable,  but   {\it persists as classical correlation  between $S$ and $T$}, since $\rho'_{ST}$ remains classically correlated. This correlation can be measured by its  mutual information  
\begin{equation}
\begin{split}I'
(S,T)
&=S(\rho'_{S})-S(S|T')\\
&=S(\rho_S)=E(S,T)\,, \label{IABT}
\end{split}
\end{equation}
since $\rho'_{S}=\rho_{S}$ while   $S(S|T')=S(\rho'_{S,T})-S(\rho'_T)=\frac{1}{N}\sum_t S(\rho_S(t))=0$.
Eq.\ \eqref{IABT} is just half 
 that of the  original reduced state \eqref{42}, and hence coincides exactly with the original $S-T$ entanglement entropy, which  then remains as {\it  classical-type mutual information}. Nevertheless, it is still  a measure of the {\it degree of evolution of the states $\rho_S(t)$}, vanishing iff all $\rho_S(t)$ are identical (i.e.\ all states $\ket{\psi_S(t)}$ identical up to a phase) and reaching its maximum $\log N$ iff all $\rho_S(t)$ are orthogonal  ($\braket{\psi_S(t)}{\psi_S(t')}=\delta_{tt'}$).  

The same result actually holds 
for {\it any} set of $N$ orthogonal rank one projectors at the clock $\tilde\Pi_j=\mathbbm{1}_S\otimes |\tilde j\rangle\langle\tilde j|$,  i.e., an unread   local  measurement in {\it any  orthogonal clock basis} of states $\{ |\tilde j\rangle,\,\langle\tilde j|\tilde j'\rangle=\delta_{jj'}\}$. We first note that relative to this basis, the history state \eqref{HS1} becomes  
\begin{equation}
    \ket{\Psi}=\sum_j \sqrt{p_j}\,|\tilde\psi_S(j)\rangle\otimes|\tilde j\rangle\,,
    \label{HSj}
\end{equation}
where 
\begin{equation}  \sqrt{p_j}|\tilde\psi_S(j)\rangle=\langle\tilde j|\Psi\rangle=\tfrac{1}{\sqrt{N}}\sum_t\langle \tilde j|t\rangle|\psi_S(t)\rangle\,,\label{condj}  \end{equation}
defines the (normalized) conditional states $|\tilde\psi(j)\rangle$ 
and the associated probabilities  $p_j=\langle \Psi|\tilde \Pi_j|\Psi\rangle$, 
$\sum_j p_j=1$. Thus, in a general basis  the history state \eqref{HS1} becomes a {\it weighted} history state.  
For   
\begin{equation}
\begin{split}\tilde\rho'_{ST}&=\tilde{\cal E}(\rho_{ST})=\sum_j  \tilde\Pi_ j\rho_{ST}\, \tilde\Pi_j\\&=\sum_jp_j\tilde\rho_{S}(j)\otimes |\tilde j\rangle\langle \tilde j|\,, \label{rhoABTtildePi} 
\end{split}\end{equation}
where $\tilde\rho_S(j)=|\tilde\psi_S(j)\rangle\langle\tilde\psi_S(j)|$, we obtain  again 
 $\tilde\rho'_{S}={\rm Tr}_T\tilde\rho'_{ST}=\sum_j p_j\tilde\rho_{S}(j)=\rho_{S}$, $\tilde\rho'_T=\sum_j p_j|\tilde j\rangle\langle\tilde j|$ 
and 
\begin{equation}\begin{split} \tilde I'(S,T)&= S(\tilde\rho'_{S})-S(S|\tilde T')\\
&=S(\rho_S)=I'(S,T)=E(S,T)\,,
\end{split}
\label{IpABT}
\end{equation}
since $S(S|\tilde T')\!=\!S(\tilde\rho'_{S,T})-S(\tilde\rho'_T)\!=\!\sum_j p_jS(\tilde \rho_S(j))=0$. 
Thus, we obtain  \\
{\bf Proposition  1.} 
{\it The system-time entanglement $E(S,T)$ of the original history state 
is the mutual information $I'(S,T)$ of the classically correlated mixed state $\tilde\rho'_{ST}$ obtained after an  unread local measurement in an arbitrary orthogonal clock basis.}  

Then, this  mutual information can likewise be used as a measure of the degree of evolution of the system $S$, now reflecting the  classical-type system-clock correlations after measurement of the history state. 
For instance, for a stationary state $|\psi_{S}(t)\rangle=e^{-i\phi(t)}|\psi_{S}(0)\rangle$,  $|\Psi\rangle=|\psi_{S}(0)\rangle|\phi_T\rangle$, with $|\phi_T\rangle=\frac{1}{\sqrt{N}}\sum_t e^{-i\phi(t)}
|t\rangle$. Then 
$\rho'_{ST}=\rho_{S}(0)\otimes \rho'_T$ (with $\rho'_T=\mathbbm 1_{T}/N$ for a measurement in the basis $\{|t\rangle\}$ and $\rho'_T=\sum_j p_j|\tilde j\rangle\langle \tilde j|$ otherwise,  with $p_j=|\langle\tilde j|\phi_T\rangle|^2$) and hence $E(S,T)=I'(S,T)=0$.

On the other hand, if the states $|\psi_S(t)\rangle$ are all orthogonal,  $|\Psi\rangle$ is  maximally entangled, 
implying $\rho'_{ST}$ having maximum classical correlation in any basis:  $p_j=|\langle \tilde j|\Psi\rangle|^2=1/N$ and $\langle\tilde \psi_S(j')|\tilde\psi_S(j)\rangle=\delta_{jj'}$ for any orthogonal clock basis, leading to 
 $I'(S,T)=\log N$.

We also notice that in a general clock basis, the relative evolution is not necessarily unitary: Eq.\ \eqref{condj} implies 
\begin{equation}
\begin{split}    \sqrt{p_j}|\tilde\psi_S(j)\rangle&=
R_j|\psi_S(0)\rangle\,,\label{Rj1}
\\
R_j&=\tfrac{1}{\sqrt{N}}\;\sum_t\langle \tilde j|t\rangle U(t)\,,
\end{split}
\end{equation}
where   the Kraus-like operators $R_j$ are non-unitary in general but satisfy  
\be \sum_j R_j^\dag R_j={\mathbbm 1}_S\,,\ee
ensuring $\sum_j p_j=1$. They lead in general to non uniform probability distributions $\{p_j\}$ in the relative history. 

\subsection{Subsystem history and evolution}

Let us now focus on the evolution of subsystem $A$ in 
\begin{equation}|\psi_S(t)\rangle\equiv|\psi_{AB}(t)\rangle=U_{AB}(t)|\psi_{AB}(0)\rangle\,,\label{ABt}\end{equation}
for a {\it general} initial state $|\psi_{AB}(0)\rangle$. 
We  may write \eqref{ABt}  as$\!\!$
\be |\psi_{AB}(t)\rangle=\sum_k \sqrt{p_k(t)}|k_A(t)\rangle |k_B\rangle\,,\label{ABtk}\ee
where $\{|k_B\rangle\}$ is any orthogonal fixed basis of $B$ ($\langle k'_B|k_B\rangle=\delta_{kk'}$) and 
\be \sqrt{p_k(t)}|k_A(t)\rangle=\langle  k_B|\psi_{AB}(t)\rangle\,,\label{e18}\ee 
such that  $|k_A(t)\rangle$ are the normalized conditional states of $A$  (not necessarily orthogonal) when $B$ is in state $|k_B\rangle$ and   $p_k(t)=\langle\psi_{AB}(t)| k_B\rangle\langle k_B|\psi_{AB}(t)\rangle$. 

If system $B$ is traced out in the history state \eqref{HS1}, we obtain a 
{\it mixture of conditional history states} of $A$, 
\begin{eqnarray}\rho_{AT}&=&{\rm Tr}_B|\Psi\rangle\langle\Psi|=\sum_k p_k|\Psi_{A_k}\rangle\langle\Psi_{A_k}|,\label{rhoAT}
\end{eqnarray}
where,  using Eq.\ \eqref{ABtk}--\eqref{e18}, 
\begin{equation}
    |\Psi_{A_k}\rangle=\tfrac{1}{\sqrt{N p_k}}\sum_{t}\sqrt{p_k(t)}|k_A(t)\rangle|t\rangle
    \,,\label{psik2}
\end{equation}
is the normalized  conditional  history state of  $A$ given that $B$ is in state $|k_B\rangle$, 
and  
$\,p_k=\tfrac{1}{N}\sum_t p_k(t)\,$ is the total  weight of state $|k_B\rangle$ in the full history ($\sum_k p_k=1$). 
 The reduced state of $A$ at time $t$ can then be recovered as 
\ben \rho_A(t)&=&{\rm Tr}_B|\psi_{AB}(t)\rangle\langle \psi_{AB}(t)|=N\langle t|\rho_{AT}|t\rangle\nonumber\\
&=&\sum_k p_k(t)|k_A(t)\rangle\langle k_A(t)|\label{rat}\,.\een

    From Eqs.\ \eqref{ABt}-\eqref{ABtk} we note that 
\begin{equation}
    \langle k_B|\psi_{AB}(t)\rangle=\sum_{k'}E_{kk'}(t)\langle k'_B|\psi_{AB}(0)\rangle\,,
\end{equation}
where 
\begin{equation}E_{kk'}(t)=\langle k_B|U_{AB}(t)|k'_B\rangle\label{Kr}\end{equation}
are {\it generalized Kraus operators} on ${\cal H}_A$, such that 
\be U_{AB}(t)=\sum_{k,k'}E_{kk'}(t)\otimes|k_B\rangle\langle k'_B|\,.\ee
 They form together  the {\it unitary  Kraus matrix of operators}$\!\!\!$ 
\begin{equation}{\bf E}(t)=\begin{pmatrix}
E_{11}(t)&\ldots&E_{1n}(t)\\
\vdots&\ddots&\vdots\\E_{n1}(t)&\ldots&E_{nn}(t)\end{pmatrix}\,,
\end{equation}
which satisfies, noting that $E_{kk'}^\dag(t)=\langle k'_B|U^\dag_{AB}(t)|k_B\rangle$, 
\be {\bf E}^\dag(t) {\bf E}(t)={\bf I}={\bf E}(t){\bf E}^\dag(t)\,,\label{Emat}\ee
i.e.\
$\sum_{l}E^\dag_{lk}(t)E_{lk'}(t)=\delta_{kk'}\mathbbm{1}_A=\sum_{l} E_{kl}(t) E_{k'l}^\dag(t)$. 
For a {\it general  entangled initial state} $|\psi_{AB}(0)\rangle=\sum_k\sqrt{p_k(0)}|k_A(0)\rangle|k_B\rangle$, we then obtain  
\begin{subequations}
\begin{align}
\rho_A(t)&=\!\!\sum_{k,k',k''}
  \!\!E_{kk'}(t)|\tilde k'_A(0)\rangle\langle\tilde k''_A(0)| E^\dag_{kk''}(t)\!\!\\&={\rm Tr}\,[{\bf E}(t)\tilde{\bm\psi}_A(0)\tilde{\bm\psi}_A^\dag(0){\bf E}^\dag(t)]\,,\label{EM}\end{align}
  \label{EMF}
  \end{subequations}
$\!\!$where $\tilde{\bm\psi}_A(0)$ denotes a column vector of elements $|\tilde k_A(0)\rangle\equiv\sqrt{p_k(0)}|k_A(0)\rangle$. For a general  {\it mixed} initial  state $\rho_{AB}(0)=\sum_\alpha q_\alpha \ketbra{\psi_{AB}^\alpha(0)}$, $q_\alpha\in[0,1]$, we should just sum \eqref{EM} over $\alpha$ with $\tilde\psi_A(0)\rightarrow \sqrt{q_\alpha}\tilde\psi_A^\alpha(0)$. 

In the case of a product initial state $|\psi_{AB}(0)\rangle=|\psi_A(0)\rangle|\psi_B(0)\rangle$, 
 Eq.\ \eqref{EM} reduces to the usual result 
\begin{equation}
    \rho_A(t)=\sum_k E_{k}(t)\rho_A(0)E^\dag_k(t)\,,\label{Eks}
\end{equation}
with $E_{k}(t)\!=\!\langle k_B|U_{AB}(t)|\psi_B(0)\rangle\!=\!\sum_{k'}\!E_{kk'}(t)\langle k'_B|\psi_B(0)\rangle$  the standard Kraus operators associated to the initial environment state $|\psi_B(0)\rangle$. If we choose $\ket{\psi_B(0)}$ to be one  element, say $|k_{0B}\rangle$, of the basis $\{|k_B\rangle\}$,  then $E_k(t)=E_{kk_0}(t)$ and the standard Kraus operators are just a column of the ${\bf E}(t)$ matrix.  

It should be emphasized, however, that for an entangled initial state, the relation of $\rho_{A}(t)$ with 
$\rho_{A}(0)$ is not through a simple state-independent map \eqref{Eks}, but rather through  the matrix ${\bf E}(t)$ applied to the vector $\tilde{\bm\psi}_A(0)$ of initial conditional components, as seen in \eqref{EMF},  since each  $|k'_A(0)\rangle$  
 is  coupled to a different environment state $|k'_B\rangle$, which determines its own set of 
maps $E_{kk'}(t)$. So far the orthogonal basis $\{|k_B\rangle\}$ is arbitrary, but could be chosen as that of the Schmidt representation of $|\Psi_{AB}(0)\rangle$, in which case the  $|\tilde k_A(0)\rangle$ are also orthogonal. 

\subsection{Quantifying subsystem evolution through mutual information}

In section B we showed that the entanglement entropy $E(S,T)$ of a pure  history state provides a consistent measure of the  ``amount'' of evolution of the whole system $S=A+B$. Moreover, after an unread 
 measurement at the clock in any orthogonal basis, this entropy remains  as the mutual information $I'(S,T)$ of the 
 ensuing classically correlated post-measurement state (Eqs.\ \eqref{IABT}-\eqref{IpABT}).  
 
 These features  no longer hold when considering just the evolution of a subsystem ($A$ or $B$) with respect to  the clock.  In the first place,  
  the subsystem reduced state, $\rho_A(t)$ isospectral with $\rho_B(t)$,  is mixed if $|\psi_{AB}(t)\rangle$ is entangled, and normally does not undergo a unitary evolution unless $U_{AB}(t)=U_A(t)U_B(t)$ $\forall\,t$. Besides, 
 a simple example below shows that the entanglement of formation \cite{W.97,W.98}  of the reduced history state $\rho_{AT}$,  Eq.\ 
 \eqref{rhoAT},  obtained after tracing out  $B$ in $|\Psi\rangle$,  cannot be directly associated to a measure of the evolution of the subsystem reduced state $\rho_A(t)$.  It also differs from $I'(A,T)$. 
 
  For instance, if $|\psi_{AB}(t)\rangle=
|\psi_A(t)\rangle|\psi_B(t)\rangle\,\forall\,t$, as occurs for an initial product state when $U(t)=U_A(t)\otimes U_B(t)$,  
the history state \eqref{HS1} becomes \footnote{we omit in what follows the obvious tensor product symbol $\otimes$ in the notation of pure composite states}  
\begin{equation}    |\Psi\rangle=\frac{1}{\sqrt{N}}\sum_t |\psi_A(t)\rangle|\psi_B(t)\rangle|t\rangle\,.\label{statep}
\end{equation}
The reduced state of $A$ at time $t$, Eq.\ \eqref{rat}, is obviously $\rho_A(t)=\ketbra{\psi_A(t)}$, independent of the states $|\psi_B(t)\rangle$. However,  $\rho_{AT}$ in Eq.\ \eqref{rhoAT} depends on the states of $B$: If e.g.\ $|\psi_{B}(t)\rangle$ evolves onto a new orthogonal state at each step, such that $\langle\psi_B(t')|\psi_B(t)\rangle=\delta_{tt'}$, 
$E(AB,T)$ is maximum, irrespective of $|\psi_A(t)\rangle$. However, after tracing out  $B$, the reduced state \eqref{rhoAT} of $AT$  
becomes 
 $\rho_{AT}=\tfrac{1}{N}\sum_t \ketbra{\psi_{A}(t)}
 \otimes \ketbra{t}$, 
 which is a {\it separable} state (i.e.\ a convex mixture of product states), hence with zero entanglement of formation: $E(A,T)=0$, irrespective of the states $\ket{\psi_A(t)}$. 
  In contrast, if $\ket{\psi_B(t)}=\ket{\psi_B(0)}$ $\,\forall\,t$ (``frozen'' system $B$), $\rho_{AT}=\ketbra{\Psi_A}$  becomes a pure history state,   
 with $\ket{\Psi_A}=\frac{1}{\sqrt{N}}\sum_t\ket{\psi_A(t)}\ket{t}$, and $E(A,T)=S(\rho_A)$, with $\rho_A=\frac{1}{N}\sum_t\ketbra{\psi_A(t)}$, which is a proper measure of the evolution of the states $\ket{\psi_A(t)}$. 
 
 Therefore, $E(A,T)$ in a composite system $ABT$,  cannot be  considered  in general a measure of the ``degree of  evolution'' of $\rho_A(t)$ with $t$.
 It is instead a measure of the quantumness of the $A-T$  correlation, absent in the first case  since the orthogonal environment states $|\psi_B(t)\rangle$ at each time prevent quantum coherence in the reduced $A+T$ subsystem  after $B$ is traced out.  Such  quantumness affects nevertheless the amount of evolution with respect to different clock basis, as discussed in next section.

However, as we now show, the subsystem-clock  mutual information $I'(A,T)$, obtained after an unread measurement in the standard clock basis $\{\ket{t}\}$ of the reduced history state $\rho_{AT}$, can still be considered as a quantifier of the evolution of the reduced state $\rho_A(t)$ along the history. 
 Tracing out $B$ in Eq. \eqref{S}, or equivalently, performing a measurement in the standard clock basis on the reduced history state \eqref{rat}, 
leads to 
\begin{equation}
    \rho'_{AT}=\tfrac{1}{N}\sum_t \rho_A(t)\otimes \ketbra{t}\,.\label{rhopat}
\end{equation}
  where $\rho_A(t)$ is now mixed in general. 
  Therefore, the  mutual information of $\rho'_{AT}$ becomes 
\begin{subequations}
\begin{align}I'(A,T)&=S(\rho'_A)+S(\rho'_T)-S(\rho'_{AT})\\
&=S(\tfrac{1}{N}\sum_t\rho_A(t))-\tfrac{1}{N}\sum_t S(\rho_{A}(t))\,,\label{IpiAT0}\\&=E(A,BT)-\bar{E}(A,B)\,,
\label{IpiAT00}
\end{align}
\label{Hol}
\end{subequations}
$\!\!$since $\rho'_A={\rm Tr}_T\rho'_{AT}=\frac{1}{N}\sum_t\rho_A(t)=\rho_A$ is the reduced state of $A$ in the whole history (i.e., the time average of $\rho_A(t)$), 
with $E(A,BT)=S(\rho_A)$ the $A-BT$ entanglement entropy in the history state, while  $S(\rho'_{AT})-S(\rho'_T)=S(A|T')
=\tfrac{1}{N}\sum_t S(\rho_{A}(t))$ is the average entropy of $\rho_A(t)$, then representing the time-average 
$\bar{E}(A,B)$ of the ``internal'' $A-B$ entanglement entropy    $E_{AB}(t)=S(\rho_{A(B)}(t))$ 
of  $|\psi_{AB}(t)\rangle$.  
Thus $I'(A,T)$ is  the ``Holevo $\chi$ quantity'' of the set $\{p(t)=\frac{1}{N},\rho_A(t)\}$ \cite{Ho.73,NC.00,schu.06}. 

 From the concavity of $S(\rho)$ (or the properties of $\chi$) it follows that $I'(A,T)$ satisfies: \\ i) $I'(A,T)\geq 0$, with $I'(A,T)=0$ iff all $\rho_{A}(t)$ are equal  ($\rho_A(t)=\rho_A$ $\forall\, t$). \\
ii) $I'(A,T)$ is maximum  when the $\rho_{A}(t)$ are all orthogonal: ${\rm Tr}\,[\rho_{A}(t)\rho_{A}(t')]=0$ $\forall\,t\neq t'$, entailing orthogonal supports  
($\rho(t)\rho(t')=0$ if  $t\neq t'$ \footnote{If $\rho(t)=\sum_\nu p_\nu(t)\ketbra{\nu(t)}{\nu(t)}$ (spectral decomposition), then ${\rm Tr}\,[\rho(t)\rho(t')]=\sum_{\nu,\nu'}|\langle \nu(t)|\nu'(t')\rangle|^2 p_\nu(t)p_{\nu'}(t')\geq 0$, vanishing iff each term vanishes, i.e.\ iff their supports are orthogonal.}), in which case 
\begin{equation} I'(A,T)=\log N\,,\label{ipatt}\end{equation}
as 
$S(\rho_A)=S(\tfrac{1}{N}\sum_t \rho_{A}(t))=
\frac{1}{N}\sum_t S(\rho_A(t))+\log N$. \\
iii) $I'(A,T)$ is then a measure of the effective number $N_{\rm eff}$ of orthogonal states $\rho_A(t)$ visited in the history, such that $N_{\rm eff}=2^{I'(A,T)}$ (for $\log=\log_2$). 
Hence, we  claim: \\
{\bf Proposition 2} {\it The mutual information \eqref{IpiAT0} provides a measure of the ``degree'' of evolution of the subsystem reduced state $\rho_A(t)$ with respect to the clock basis $\{|t\rangle\}$, quantifying the effective number of orthogonal  states $\rho_A(t)$ visited through the history.}

On the other hand, this mutual information now differs from the subsystem-clock entanglement $E(A,T)$ in the reduced history state \eqref{rhoAT}.  
For instance, if we consider again the product case of Eq.\ \eqref{statep}, in which $E(A,T)$ ranges from $0$ (orthogonal states $|\psi_B(t)\rangle$) to $S(\rho_A)$ (freezed $|\psi_B(t)\rangle=|\psi_B(0)\rangle$), we
 now obtain 
\begin{equation}I'(A,T)=S(\rho_A)\label{34}\,,\end{equation} 
irrespective of the states $|\psi_B(t)\rangle$, 
since here $\rho_A(t)=\ketbra{\psi_A(t)}$  $\forall\,t$ and hence 
the second term in \eqref{IpiAT0}  vanishes. Eq.\ \eqref{34} is the expected result 
for this case, which would come out as standard system-time entanglement entropy of $A$ alone if $B+T$ were considered 
in \eqref{statep} as an extended clock with orthogonal states $|\psi_{B}(t)\rangle |t\rangle$, 
or if system $B$ would be ignored. 

 Note also that  since the last term in \eqref{IpiAT00} is the same for $A$ and $B$, as $\rho_A(t)$ is  isospectral to $\rho_B(t)$,  \eqref{Hol} implies
\be I'(A,T)-I'(B,T)=S(\rho_A)-S(\rho_B)\,,\label{dif}\ee
which can be non-zero ($\rho_{A(B)}=\frac{1}{N}\sum_t \rho_{A(B)}(t)$). 

Finally, if $|\psi_{AB}(t)\rangle=U_A(t)\otimes U_B(t)|\psi_{AB}(0)\rangle\, \forall\,t$, i.e.\ $A-B$ noninteracting,  then $\rho_{A}(t)=U_A(t)\rho_A(0)U_A^\dag(t)$ for any initial state, undergoing a unitary evolution. Hence  $S(\rho_A(t))=S(\rho_A(0))$ and $I'(A,T)$ depends only on $U_A(t)$ and the time-independent $A-B$ entanglement entropy: 
\begin{equation}
\begin{split}
    I'(A,T)&=E(A,BT)-E(A,B)=S(\rho_A)-S(\rho_A(0))
\end{split}
\end{equation}
\vspace*{-0.35cm}

\subsection{Subsystem evolution relative to a general clock basis}\label{II.E}
\vspace*{-0.35cm}

The history state formalism allows in principle to consider different clock bases for describing the relative evolution of a subsystem. And in contrast with the whole system $S=A+B$, where the postmeasurement  mutual information $I'(S,T)$ is independent of the choice of the clock basis, always coinciding with the history entanglement entropy $E(S,T)$, the degree of evolution of a subsystem with respect to $T$ will {\it depend} in general on the clock basis. This implies that there is an `optimum' basis for which the relative subsystem evolution is maximal.  

Let us then consider a general orthogonal basis $\{\ket{\tilde j}\}$ of $T$. After performing a measurement in this basis and tracing out the ``environment'' $B$ in Eq.\ \eqref{rhoABTtildePi}, we obtain 
\begin{equation}\tilde\rho'_{AT}=\sum_{j}p_j\tilde\rho_A(j)\otimes |\tilde j\rangle\langle \tilde j|,\label{rhoATtilde}
\end{equation}
where $\tilde\rho_A(j)={\rm Tr}_B\,\tilde\rho_{AB}(j)$. Its mutual information is 
\begin{subequations}
\begin{align}
\tilde I'(A,T)
&=S(\tilde\rho'_A)-S(A|\tilde T')\rangle \label{ItpAT1}
\\
&=S(\rho_A)-\sum_j p_jS(\tilde\rho_{A}(j))\,,\label{ipj2}
\end{align}
\label{ipjg}
\end{subequations}
\vspace*{-0.35cm}

$\!\!\!$since $\tilde\rho'_A={\rm Tr}_T\tilde\rho'_{AT}=\sum_j p_j\tilde\rho_A(j)=\rho_A$ and $S(A|\tilde T')=S(\tilde\rho'_{AT})-S(\tilde\rho'_T)=\sum_j p_j S(\tilde\rho_{A}(j))$. It is again a 
``Holevo $\chi$ quantity'' for the set $\{p_j,\tilde\rho_A(j)\}$, with  the sum in  \eqref{ipj2}  representing again  the average $A-B$ entanglement entropy of states $|\tilde\psi_{AB}(j)\rangle$ with weights $p_j$.   

The mutual information \eqref{ipj2} provides a weighted measure of the degree of evolution of the conditional subsystem states $\tilde\rho_{A}(j)$ with respect to the  basis $\{|j\rangle\}$. For {\it any} such basis, it satisfies: \\
i) $\tilde I'(A,T)\geq 0$, with  $\tilde I'(A,T)=0$ iff  all $\tilde\rho_{A}(j)$ with $p_j>0$ are coincident. \\
ii) For fixed $\{p_j\}$, $\tilde I'(A,T)$ is maximum  when all $\tilde\rho_{A}(j)$ with $p_j>0$ are orthogonal, ${\rm Tr}\,[\tilde\rho_A(j)\tilde\rho_A(j')]=0$ $\forall\,j\neq j'$, in which case 
\begin{equation}\tilde I'(A,T)=-\sum_j p_j\log p_j\leq \log N\,.\end{equation}
iii) $\tilde I'(A,T)\leq S(\rho_A)$, 
as follows from \eqref{ipj2},
this upper bound reached when all conditional states $\tilde \rho_A(j)$ with $p_j>0$ are pure.

While Eq.\ \eqref{ipjg} will depend in general on the chosen clock basis (see next subsections), 
 the difference 
\eqref{dif} stays unchanged and is hence independent of the latter, since  
$\tilde\rho_A(j)$ remains  isospectral with $\tilde\rho_B(j)$ as they both stem from $|\tilde\psi_{AB}(j)\rangle$ (Eq.\ \eqref{condj}):   
\be \tilde I'(A,T)-\tilde I'(B,T)=S(\rho_A)-S(\rho_B)\,.\label{dif2}\ee

{\it Comparison with global evolution}.
It is apparent that the degree of evolution of a subsystem should not exceed that of the full system.  
For instance, if all $|\psi_{AB}(t)\rangle$ are maximally entangled and orthogonal,  $E(AB,T)$ is maximal, i.e.\ there is maximal full system evolution, whereas $\rho_A(t)$ will be maximally mixed  $\forall\,t$   entailing null subsystem evolution   $I'(A,T)=0$, as $\rho'_{AT}$ becomes a product state. In the general case we have the general inequality:\\ 
{\bf Proposition 3}. {\it The degree of subsystem evolution as quantified by the mutual information \eqref{IpiAT0} or in general 
\eqref{ipj2}, cannot exceed that of the  full system evolution as quantified by the entanglement entropy \eqref{e4} or equivalently, the mutual information \eqref{IpABT}: 
\begin{equation}
    \tilde I'(A,T) \leq E(AB,T)=S(\rho_{AB})\,,
  \label{desg}
\end{equation}
for ${\it any}$ local clock basis defined by projectors  $\{\tilde\Pi_j\}$.} 

This is an expected property of a consistent measure of subsystem evolution. 

{\it Proof}. Using the subadditivity property of the von Neuman entropy \cite{NC.00}, $S(\rho_A)\leq S(\rho_{AB})+S(\rho_{AT})-S(\rho_{ABT})$, we obtain, for $\rho'_{ABT}=\sum_j p_j\tilde\rho_{AB}(j)\otimes|\tilde j\rangle\langle \tilde j|$,  
$\rho'_{AB}=\rho_{AB}$, $\rho'_A=\rho_A$ and $\rho'_{AT}=\sum_j p_j\tilde\rho_A(j)\otimes |\tilde j\rangle\langle\tilde j|$, 
\begin{equation}\begin{split}
   S(\rho_{AB})&\geq S(\rho_A)-S(\rho'_{AT})+S(\rho'_{ABT})\\
    &=S(\rho_A)-\sum_j p_jS(\tilde \rho_A(j))=\tilde I'(A,T)
\end{split}\label{Iproof}\,,\end{equation}
since here $\tilde\rho_{AB}(j)=|\tilde\psi_{AB}(j)\rangle\langle\tilde\psi_{AB}(j)|$ is pure and hence 
$S(\rho'_{ABT})-S(\rho'_{AT})=\sum_j p_j S(\tilde \rho_A(j))$. 
This leads to Eq.\ \eqref{desg} since $S(\rho_{AB})=E(AB,T)$. $\qed$ 

Eq.\ \eqref{desg} is in agreement with the property that tracing out a subsystem does not increase the mutual information, as $E(AB,T)=\tilde I'(AB,T)$ (Eq.\ \eqref{IpABT}). 
\vspace*{-0.35cm}

\subsection{Maximal subsystem evolution}
\vspace*{-0.35cm}

The mutual information \eqref{ipjg} depends in general on the choice of clock basis, since both $\tilde \rho_{A}(j)$ and $p_j$ are basis dependent. The degree of basis dependence increases as the quantumness of the correlations in the reduced state $\rho_{AT}$  decreases. For instance, in the example of Eq.\ \eqref{statep}, if the environment $B$ is frozen,  $|\psi_B(t)\rangle=|\psi_B\rangle$ $\forall\,t$, $\rho_{AT}={\rm Tr}_{B}\,\ketbra{\Psi}$ becomes a standard pure history state $|\Psi_{AT}\rangle\langle\Psi_{AT}|$,   having then pure quantum correlations if subsystem $A$ is evolving. Then  Eq.\ \eqref{ipjg}
 becomes basis independent, i.e., 
 \begin{equation} 
 \tilde I'(A,T)=I'(A,T)=S(\rho_A)=E(A,T) ,\label{EATp}
 \end{equation}
as all $\tilde\rho'_A(j)$ are pure and the second term in \eqref{ipjg} vanishes.

 In contrast, if the states $|\psi_B(t)\rangle$ in \eqref{statep}  are all orthogonal, $\rho_{AT}$ becomes already  the  {\it classically correlated} state \eqref{rhopat}, remaining invariant    under a measurement in the standard clock basis only. The post-measurement mutual information \eqref{ipj2} becomes then strongly dependent on the choice of basis: While in the standard clock basis we  have $I'(A,T)=S(\rho_A)$, as previously stated, if the clock were measured in the conjugate Fourier basis $|\tilde j\rangle=\frac{1}{\sqrt{N}}\sum_t e^{i2\pi jt/N}|t\rangle$, then  $|\langle \tilde j|t\rangle|=\frac{1}{\sqrt{N}}\forall\,t$ and $p_j=\frac{1}{N}$,  
 $\tilde\rho_A(j)=\frac{1}{N}\sum_t\rho_A(t)=\rho_A$ $\forall\,j$,  obtaining instead 
\be \tilde\rho'_{AT}=\tfrac{1}{N}\rho_A\otimes \sum_j|\tilde j\rangle\langle \tilde j|\,,\label{rhpat}
\ee 
which is a {\it product} state leading to $\tilde I'(A,T)=0$. Here $\rho_A$ is the reduced state of $A$ in the full history. 
Thus, subsystem $A$ {\it is not evolving} with respect to the Fourier clock basis when $\rho_{AT}$ has already  the classically correlated form \eqref{rhopat}. 
Previous effect can also be seen  from the corresponding conditional $AB$ states \eqref{condj}, which in the case \eqref{statep} become  the {\it entangled} states 
$|\tilde\psi_{AB}(j)\rangle=\frac{1}{\sqrt{N}}\sum_t e^{-i2\pi jt/N}|\psi_A(t)\rangle|\psi_B(t)\rangle$. These states lead all to {\it the  same reduced state} 
$\tilde\rho_A(j)={\rm Tr}_B\tilde\rho_{AB}(j)=\frac{1}{N}\sum_t |\psi_A(t)\rangle\langle\psi_A(t)|=\rho_A\,\forall\,j$, when the states $\ket{\psi_B(t)}$ are all orthogonal, such that the $\tilde\rho_A(j)$ are not evolving with respect to this clock basis, even though $S=A+B$ maintains maximum evolution ($E(S,T)=\log N$ as still $\langle\tilde\psi_{AB}(j)|\tilde\psi_{AB}(j')\rangle=\delta_{jj'}$).    Thus, in this case the choice of clock basis can tune $\tilde I'(A,T)$ from $S(\rho_A)$ in the $\{\ket{t}\}$ basis to $0$ in the associated Fourier 
basis $\{|\tilde j\rangle\}$.

Therefore,  in general we have a clock-basis dependent measure of evolution of a subsystem based on the mutual information $\tilde I'(A,T)$ after a local measurement at the clock. Its  maximization  over all measurements on the clock  then leads to a basis independent measure  identifying the maximum degree of evolution achievable.
This maximization is related to the Quantum Discord \cite{OZ.01,HV.01}. 

 Given a bipartite state $\rho_{AT}$ (generally mixed) and a projective measurement $\tilde\Pi$ on the clock based on 
 projectors $\tilde\Pi_j$, the Quantum Discord  can be expressed as 
 \begin{subequations}\begin{align}
 \label{DAT}
 \hspace*{-0.25cm}D(A/T)
 &= I(A,T)-\underset{\tilde\Pi}{\rm Max}\,\tilde I'(A,T)  \\
 &=\underset{\tilde\Pi}{\rm Min}\,S(A|\tilde T')-S(A|T)\,,\label{DATB}
 \end{align}
 \end{subequations}
  where $I(A,T)=S(\rho_A)-S(A|T)$ is the original $AT$ mutual information, with $S(A|T)=S(\rho_{AT})-S(\rho_T)$ the quantum conditional entropy, while  $\tilde I'(A,T)=S(\rho_A)-S(A|\tilde T')$, with $S(A|\tilde T')=\sum_j p_jS(\rho_A(j))$ that  after an unread local measurement $\{\tilde\Pi\}$  on the clock $T$. 

{\bf Proposition 4}. {\it  The maximum subsystem-time  mutual information obtained after a measurement on the clock, is the difference between the total entanglement $E(A,BT)$ between $A$ and the remaining system, and the internal system entanglement $E(A,B)$:} 
\begin{equation}
 \underset{\tilde\Pi}{\rm Max}\, \tilde{I'}(A,T)=E(A,BT)-E(A,B).
\label{lem3}
\end{equation}

{\it Proof}. From Eq.\ \eqref{DATB}, using that the minimum of the conditional entropy $S(A|\tilde T')$ over  the whole set of 
local measurements $\{\tilde\Pi\}$ on the clock 
is the entanglement of formation $E(A,B)$ between $A$ and the system $B$ which purifies the whole system \cite{KW.04,GR.13}, and that $E(A,BT)=S(\rho_A)=S(\rho_{BT})$ is the total entanglement between $A$ and the remaining environment-clock system,  
we obtain 
\begin{equation}
\begin{split}
 \underset{\{\tilde\Pi\}}{\rm Max}\, \tilde I'(A,T)&= I(A,T)-D(A/T)  \\
 &=S(\rho_A)-\underset{\{\tilde\Pi\}}{\rm Min}\, S(A|T')
\end{split}\end{equation}
 which leads to \eqref{lem3}. 
Thus, the maximum $\tilde I'(A,T)$  is essentially the remnant $A-T$ entanglement after subtracting the entanglement of formation $E(A,B)$ of $\rho_{AB}$  from the total  $A-BT$ entanglement. $\qed$ 

In contrast, ${\rm Min}_{\{\tilde\Pi\}}\tilde I'(A,T)=E(A,BT)-E_a(A,B)$, where 
$E_a(A,B)={\rm Max}_{\{\tilde\Pi\}}\, S(A|T')$ is 
the entanglement of assistance \cite{Cohen.98, DiV.99}.

\section{Illustrative examples}\label{sec:examples}

\subsection{A Qubit Clock}
We first consider for illustration a two-times history of a general bipartite system $A+B$, corresponding to the 
  case of a qubit clock $T$, 
\begin{equation}
|\Psi\rangle=\tfrac{1}{\sqrt{2}}(|\psi_{AB}^0\rangle|0\rangle+|\psi_{AB}^1\rangle|1\rangle)\,.\label{psiq2}\end{equation}
The full system--time entanglement entropy 
is fully  determined by the overlap between both system states: 
\begin{equation}
\!\!\! E(AB,T)=-\sum_{\nu=\pm}\lambda_\nu \log\lambda_\nu\,,\;
\lambda_{\pm}=\tfrac{1\pm|\braket{\psi_{AB}^0}{\psi_{AB}^1}|}{2}
\label{EF}
\end{equation}
where $\lambda_{\pm}$ are the eigenvalues of the reduced states $\rho_{AB}=\frac{1}{2}\sum_t\ketbra{\psi_{AB}^t}{\psi_{AB}^t}$ and $\rho_T$. 
This entanglement is a decreasing function of $|\langle\psi_{AB}^0|\psi_{AB}^1|$, reaching its maximum $1$  for orthogonal states, i.e., maximal evolution, and vanishing when they are equal up to a phase.  

It also coincides with the whole system--time mutual information obtained after a  measurement in {\it any} clock basis. In the computational basis 
$\{|0\rangle,|1\rangle\}$, the unread post measurement state is just
\begin{equation}    \rho'_{ABT}=\frac{1}{2}\sum_{t=0,1}|\psi_{AB}^t\rangle\langle\psi_{AB}^t|\otimes |t\rangle\langle t|\,,\label{rabt}
\end{equation}
and its system-time  mutual information is clearly  $I'(AB,T)=S(\rho_{AB})=E(AB,T)$.  
In an orthogonal clock basis 
$B_\alpha=\{\ket{\alpha_+},\ket{\alpha_-}\}$, with 
\begin{equation}
\begin{split}
\ket{\alpha_+}&=\cos\tfrac{\alpha}{2}\ket{0}+\sin\tfrac{\alpha}{2}\ket{1}\\\ket{\alpha_-}&=\sin\tfrac{\alpha}{2}\ket{0}-\cos\tfrac{\alpha}{2}\ket{1}\end{split}\,,\label{sal}
\end{equation} 
Eq.\ \eqref{psiq2} is rewritten as a weighted history state 
\begin{equation}
\ket{\Psi}=\sqrt{p^\alpha_+}\ket{\psi_{AB}^{\alpha_+}}|\alpha_+\rangle+\sqrt{p^\alpha_-}\ket{\psi_{AB}^{\alpha_-}}\ket{\alpha_-}\,,
\label{Psial}
\end{equation}
where $\ket{\psi_{AB}^{\alpha_{\pm}}}$ are the  normalized states 
\begin{equation}
\label{psial}
\begin{split}
\ket{\psi_{AB}^{\alpha_+}}&=\tfrac{1}{\sqrt{2p^\alpha_+}}(\cos\tfrac{\alpha}{2}\ket{\psi_{AB}^0}+\sin\tfrac{\alpha}{2}\ket{\psi_{AB}^1})\\ \ket{\psi_{AB}^{\alpha_-}}&=
\tfrac{1}{\sqrt{2 p_-^\alpha}}(\sin\tfrac{\alpha}{2}\ket{\psi_{AB}^0}-\cos\tfrac{\alpha}{2}\ket{\psi_{AB}^1})\end{split}\end{equation} 
and $p^\alpha_\pm=\frac{1\pm |\braket{\psi_{AB}^0}{\psi_{AB}^1}|\sin\alpha}{2}$  (we have assumed  a real overlap without loss of generality \footnote{if $\braket{\psi_{AB}^0}{\psi_{AB}^1}=e^{-i\phi}|\braket{\psi_{AB}^0}{\psi_{AB}^1}|$, the phase can be absorbed in the history state  replacing 
$\ket{1}\rightarrow e^{i\phi}\ket{1}$.}). In particular, for $\alpha=\pi/2$ the states \eqref{psial} become orthogonal and \eqref{Psial} becomes the Schmidt decomposition of \eqref{psiq2},  with $p^{\pi/2}_{\pm}=\lambda_{\pm}$ the eigenvalues of $\rho_{AB}$, $\rho_T$.  

After an unread local measurement at the clock in the general basis \eqref{sal}, the whole state becomes 
\begin{equation}    \rho^{\prime\alpha}_{ABT}=\sum_{\nu=\pm}p_\nu^\alpha|\psi_{AB}^{\alpha_\nu}\rangle\langle\psi_{AB}^{\alpha_\nu}|\otimes |\alpha_\nu\rangle\langle \alpha_\nu|\,, 
\label{rabtp}
\end{equation}
but its total system-time mutual information 
is independent of $\alpha$, as the conditional states remain pure and still 
$\sum_{\nu=\pm}p^\alpha_\nu\ketbra{\psi^{\alpha_\nu}_{AB}}{\psi^{\alpha_\nu}_{AB}}=\rho_{AB}$, implying again
$I^{\prime\alpha}_{AB,T}=S(\rho_{AB})=E(AB,T)$. 

However, the subsystem evolution with respect to a given  clock basis will depend, in general, on the clock basis: After tracing out $B$ in Eqs.\ \eqref{rabt}-\eqref{rabtp},  we obtain   
 \begin{equation}    \rho'_{AT}=\tfrac{1}{2}(\rho_A^{0}\otimes \ketbra{0}{0}+\rho_A^1\ketbra{1}{1})\,,\label{rhopatp}
\end{equation}
in the standard basis, with  $\rho_A^t={\rm Tr}_B\ketbra{\psi_{AB}^t}{\psi_{AB}^t}$, and 
\begin{equation}    \rho^{\prime\alpha}_{AT}=p^\alpha_+\rho_A^{\alpha_+}\!\otimes \ketbra{\alpha_+}{\alpha_+}+p^\alpha_-\rho_A^{\alpha_-}\!\otimes\ketbra{\alpha_-}{\alpha_-}\,,\label{rhopatal}
\end{equation}
in the basis \eqref{sal},  with $\rho_A^{\alpha_\pm}={\rm Tr}_B\ketbra{\psi_{AB}^{\alpha_{\pm}}}$, which leads in general to an $\alpha$-dependent mutual information $I^{\prime\alpha}(A,T)$, even though 
$\sum_{\nu=\pm}p^\alpha_\nu\rho^{\alpha_\nu}_A=\frac{1}{2}\sum_t\rho_A^t=\rho_A$.

\subsubsection{Generalized GHZ-type history states} As first explicit example, let us assume that both system states are product states $\ket{\psi_{AB}^t}=\ket{\psi_A^t}\ket{\psi_B^t}$ for $t=0,1$, such that 
\begin{subequations}
\begin{align}\ket{\Psi}&=\tfrac{1}{\sqrt{2}}
(|\psi_A^0\rangle|\psi_B^0\rangle|0\rangle+|\psi_A^1\rangle|\psi_B^1\rangle|1\rangle)\label{psiqs1}\\&\equiv\tfrac{1}{\sqrt{2}}(\ket{000}+\ket{\theta\gamma 1})\,.\label{psiqs2}\end{align}
\label{psitg}
\end{subequations}
Since just  two states are involved in each subsystem, the  state \eqref{psiqs1} is equivalent to a two-step two-qubit history state \eqref{psiqs2}, i.e., a generalized GHZ-type three-qubit state, where $|\theta\rangle=\cos\theta|0\rangle+\sin\theta|1\rangle$ and $|\gamma\rangle=\cos\gamma|0\rangle+\sin\gamma|1\rangle$   are the final states of $A$ and $B$, with $\cos\theta=
\langle\psi_A^0|\psi_A^1\rangle$, $\cos\gamma
=\langle\psi_B^0|\psi_B^1\rangle$ and $\langle 1|0\rangle=0$. It corresponds to a product step evolution operator $U_{AB}=U_A\otimes U_B$.  The whole system-time entanglement entropy is given by \eqref{EF} with $\lambda_{\pm}=\frac{1\pm\cos\theta\cos\gamma}{2}$.

We first notice that the reduced state of $AB$ in $|\Psi\rangle$, 
\begin{equation}
    \rho_{AB}={\rm Tr}_T\ketbra{\Psi}=\tfrac{1}{2}(\ketbra{00}+\ketbra{\theta\gamma})\,,\label{rhAB}
\end{equation}
is {\it separable} and hence $E(A,B)=0$. This entails that the maximum achievable value \eqref{lem3} of the post-measurement subsystem $A$-time  mutual information  becomes 
\begin{equation}
\!\!I_{\rm max}'(A,T)=E(A,BT)=S(\rho_A)=-
\!\!\sum_{\nu=\pm}\!\lambda^\theta_\nu
\log\lambda^\theta_\nu\,,
  \label{ipat2} 
\end{equation}
with $\lambda^\theta_\pm=\tfrac{1}{2}(1\pm\cos\theta)$
 the eigenvalues of  
\begin{equation}\rho_A={\rm Tr}_{BT}\ketbra{\Psi}=\tfrac{1}{2}(\ketbra{0}+\ketbra{\theta})\,.\label{rhaa}\end{equation}
This maximum is here clearly reached for a measurement in the {\it standard clock basis}, where the reduced $AT$ postmeasurement state  becomes 
\begin{equation}
    \rho'_{AT}=\tfrac{1}{2}(\ketbra{00}{00}+\ketbra{\theta 1}{\theta 1})\,,\label{rhopat22}
\end{equation}
implying pure conditional reduced states $\rho_A^{0}=\ketbra{0}$, $\rho_A^{1}=\ketbra{\theta}$,   
which lead to $I'(A,T)=I'_{\rm max}(A,T)$.  
  
  Hence, with respect to this basis subsystem $B$ has no effect in the evolution of $A$, such that the associated mutual information \eqref{ipat2} coincides  with the  entanglement entropy  of a pure $AT$ history state $|\Psi_{AT}\rangle=\frac{1}{\sqrt{2}}(|00\rangle+|\theta 1\rangle)$. This is apparent as $BT$ states $|00\rangle$ and $|\gamma 1\rangle$ in \eqref{psiqs2} can  be considered as orthogonal states of an enlarged clock.  Similar considerations hold for subsystem $B$ ($\theta\longleftrightarrow\gamma$).

However, the evolution  of $A$ and $B$  with respect to a general clock basis \eqref{sal} can change considerably, since the conditional states \eqref{psial} become in general {\it entangled}. Thus, they lead to {\it mixed} reduced states $\rho_A^{\alpha_{\pm}}$ in \eqref{rhopatal}, entailing 
a reduced distinguishability between them. 

{\it Case $\theta=\gamma=\pi/2$}.  
Let us first consider the case where  both subsystems evolve to an orthogonal state  in the standard clock basis,   such that $|\theta\rangle=|\gamma\rangle=|1\rangle$ and Eq.\ \eqref{psiqs2} becomes a standard GHZ state 
\begin{equation}
|\Psi\rangle_{\rm GHZ}
= \tfrac{1}{\sqrt{2}}(|000\rangle + |111\rangle),\label{GHZ1}
\end{equation}
corresponding to maximum subsystem evolution $I'(A,T)=I'(B,T)=1$. 
However, a change to the conjugate  ``momentum'' clock basis $|\pm\rangle=\tfrac{|0\rangle\pm |1\rangle}{\sqrt{2}}$ ($\alpha=\pi/2$ in \eqref{sal}) 
leads to {\it maximally entangled} orthogonal $AB$  states \eqref{psial} \cite{Popp.04}  at each new ``time'' $|\pm\rangle$: 
 \begin{equation}\ket{\Psi}_{\rm GHZ}= 
    \tfrac{1}{\sqrt{2}}(\tfrac{|00\rangle+|11\rangle}{\sqrt{2}}|+\rangle+\tfrac{|00\rangle- |11\rangle|}{\sqrt{2}}|-\rangle).\label{GHZ2}
\end{equation}
Thus, with respect to this clock basis, while the whole system  again evolves between orthogonal states, in agreement with the invariant total entanglement $E(AB,T)$,  reduced subsystem states {\it are not evolving}, 
 staying instead maximally mixed: $\rho^{\prime \pi/2_{\pm}}_{A(B)}=\frac{1}{2}\mathbb 1$ and hence
$\rho^{\prime\pi/2}_{AT}=\rho^{\prime\pi/2}_{BT}=\tfrac{1}{4}\mathbbm{1}\otimes(\ketbra{+}{+}+\ketbra{-}{-})=\tfrac{1}{4}\mathbbm{1}\otimes\mathbbm{1}$, leading accordingly to $I^{\prime\pi/2}(A,T)=I^{\prime\pi/2}(B,T)=0$. 

\begin{figure}
\centering
\includegraphics[scale=.9]{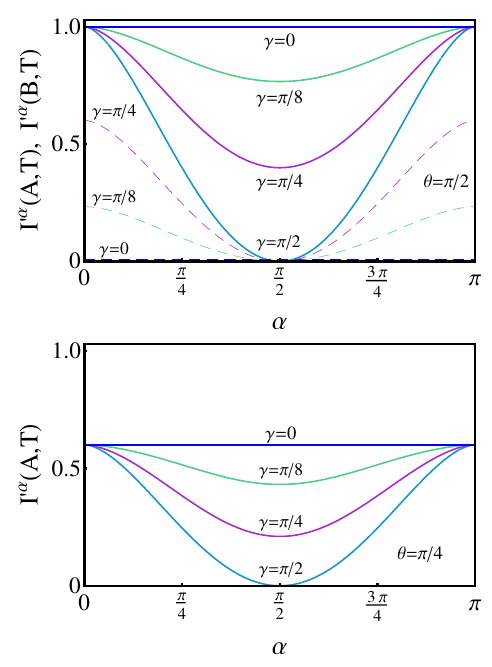}
\vspace*{-0.35cm}

\caption{Mutual information $I'^{\alpha}(A,T)$ (solid lines) and $I'^{\alpha}(B,T)$ (dashed lines, top panel), quantifying the  subsystem evolution with respect to clock basis $B_\alpha$, Eq.\ \eqref{sal}, for the history state \eqref{psitg} with  $\theta=\pi/2$ (top) and $\pi/4$ (bottom), at distinct values of $\gamma$. As the entanglement between $A$ and $T$ decreases (increasing $\gamma$), the dependence of $I^{\prime\alpha}(A,T)$ with the clock basis $B_\alpha$ increases. A similar behavior holds for $I'^{\alpha}(B,T)$ with 
$\theta\longleftrightarrow\gamma$. The difference $I'^{\alpha}(B,T)\!-\!I'^{\alpha}(A,T)$ is $\alpha$-independent (Eq.\ \eqref{dif2}).\vspace*{-0.5cm}}
\label{fig1}
\end{figure}

In fact, for general $\alpha$ in  \eqref{sal},  Eq.\ \eqref{GHZ1} can be recast as 
 \begin{equation}
 \begin{split}\ket{\Psi}_{\rm GHZ}&= 
    \tfrac{1}{\sqrt{2}}(|\psi^{\alpha_+}_{AB}\rangle|\alpha_+\rangle+|\psi^{\alpha_-}_{AB}\rangle|\alpha_-\rangle)\,,\\    |\psi^{\alpha_{\pm}}_{AB}\rangle&=\;{^{\cos\frac{\alpha}{2}}_{\sin\frac{\alpha}{2}}}|00\rangle\pm{^{\sin\frac{\alpha}{2}}_{\cos\frac{\alpha}{2}}}|11\rangle\,,\label{GHZa}\end{split}\end{equation}
where the nonmaximally entangled states $|\psi^{\alpha_{\pm}}_{AB}\rangle$ are still orthogonal but lead now to distinct yet nonorthogonal and isospectral reduced  states 
\begin{equation}
\label{psial3}
\rho_{A}^{\prime\alpha_{\pm}}={^{\cos^2\frac{\alpha}{2}}_{\sin^2\frac{\alpha}{2}}}\ketbra{0}{0}+{^{\sin^2\frac{\alpha}{2}}_{\cos^2\frac{\alpha}{2}}}\ketbra{1}{1}\,, \end{equation} 
with eigenvalues $\lambda^\alpha_{\pm}=\frac{1}{2}(1\pm \cos\alpha)$. Eq.\ \eqref{ipjg} then implies 
\begin{equation}
I^{\prime\alpha}(A,T) = 
1+ \sum_{\nu=\pm} \lambda^\alpha_\nu\,\log \lambda^\alpha_\nu\,,\label{IGHZ}
\end{equation}
which decreases from $1$ to $0$ as $\alpha$ increases from $0$ to $\pi/2$, as seen in the top panel of Fig.\ \ref{fig1} for $\gamma=\pi/2$.

{\it General $\theta$, $\gamma=\pi/2$}. Previous behavior holds in fact for any $\theta\neq 0$ in \eqref{psiqs2} if $\gamma=\pi/2$, as seen in the bottom panel for $\theta=\pi/4$, except that subsystem $A$ does not evolve now to an orthogonal state in the standard basis and hence starts from a lower value at $\alpha=0$. Eqs.\ \eqref{GHZa}-\eqref{psial3} remain valid replacing $|11\rangle\rightarrow |\theta 1\rangle$ and $\ketbra{1}{1}\rightarrow\ketbra{\theta}{\theta}$, while \eqref{IGHZ}
is replaced by 
\begin{subequations}
\begin{align}
I^{\prime\alpha}(A,T)&=S(\rho_A)-S(\rho_A^{\prime\alpha\pm})\\
&=-\sum_{\nu=\pm}(\lambda^\theta_\nu\log\lambda^\theta_\nu-\lambda^{\alpha\theta}_\nu\log\lambda^{\alpha\theta}_\nu)
\label{IGHZpal22}\,,\end{align}
\end{subequations}
where $\lambda_\nu^\theta=\lambda_\nu^{\alpha=\frac{\pi}{2},\theta}$ and \begin{equation}    \lambda^{\alpha\theta}_{\pm}=\tfrac{1}{2}
    (1\pm\sqrt{1-\sin^2\alpha\sin^2\theta})\,,
\end{equation}
are the eigenvalues of both $\rho^{\prime\alpha_{\pm}}$, which remain isospectral $\forall\alpha$ and coincident at $\alpha=\pi/2$, thus leading again to $I^{\prime\alpha}(A,T)=0$ at $\alpha=\pi/2$.

 {\it General $\theta,\gamma$}.
From the perspective of  the $AT$ subsystem, the role of the angle $\gamma$ between the initial and final states of $B$  is to control the quantum correlations between $A$ and $T$ in the reduced history state $\rho_{AT}={\rm Tr}_B\ketbra{\Psi}$, 
 \begin{equation}
 \begin{split}
 \rho_{AT}=&\tfrac{1}{2}[\ketbra{00}+\ketbra{\theta 1}\\&+\cos\gamma(\ketbra{00}{\theta 1}+\ketbra{\theta 1}{00})].\label{ratx}
 \end{split}
 \end{equation}
 It is in general a mixed state (isospectral with $\rho_B$) with eigenvalues $\lambda^\gamma_{\pm}=\frac{1}{2}(1\pm\cos\gamma)$  and concurrence \cite{W.97,W.98} 
\begin{equation}
    C_{AT}=|\sin\theta\cos\gamma|\,,
    \label{Cat}
\end{equation}
hence {\it entangled} for $\gamma\neq \pi/2$ (if $\theta\neq 0$). Its entanglement of formation $E(A,T)=-\sum_{\nu=\pm}\lambda^C_\nu\log \lambda^C_\nu$, with 
$\lambda^C_\pm=\frac{1\pm\sqrt{1-C_{AT}^2}}{2}$, then satisfies  $E(A,T)\leq I'(A,T)$, Eq.\ \eqref{ipat2}, coinciding just in the trivial cases $\gamma=0$ or $\theta=0$.

The previous strong variation of $I^{\prime\alpha}(A,T)$ with $\alpha$ observed for $\gamma=\pi/2$, stems in fact from the {\it separability} of the reduced state \eqref{ratx} at this $\gamma$, for which $E(A,T)=C_{AT}=0$ and \eqref{ratx} becomes, moreover,  {\it classically correlated}. Thus, it leads to pronounced clock-basis dependence of $AT$ correlations after measurement,  which vanish in the conjugate Fourier basis (here states $|\pm\rangle$, $\alpha=\pi/2$), according to the discussion below Eq.\ \eqref{rhpat}.

In contrast, for $\gamma\neq \pi/2$, $\rho_{AT}$ becomes entangled if $\theta\neq 0$, and this  entails a softening of the $\alpha$-dependence of $I^{\prime\alpha}(A,T)$, {\it preventing it from vanishing} at any $\alpha$.  In fact, it is shown in  Appendix B that if there is no subsystem evolution for some clock basis (e.g.\ $I^{\prime\alpha}(A,T)=0$ for some $\alpha$) in a  general two-qubit two-times scenario,  the original reduced state $\rho_{AT}$ {\it is necessarily separable} ($C_{AT}=0$). 

 For general $\gamma$, after a clock measurement  in the basis \eqref{sal}, the conditional $AB$ states \eqref{psial} lead  to $\gamma$-dependent reduced states of $A$, 
 \begin{equation}
\label{psial4}
\rho_{A}^{\alpha_\pm}\propto\tfrac{1\pm\cos\alpha}{2}\ketbra{0}+\tfrac{1\mp\cos\alpha}{2}\ketbra{\theta}\pm\sin\alpha\cos\gamma\tfrac{\ketbra{0}{\theta}+\ketbra{\theta}{0}}{2},
\end{equation} 
 with probabilities $p^\alpha_{\pm}=\tfrac{1}{2}(1\pm\sin\alpha\cos\theta\cos\gamma)$. 
 These states are now distinct  and not isospectral if  
 $\sin\alpha\cos\gamma\neq 0$, having (after normalization) eigenvalues 
  \begin{equation}     \lambda^{\alpha_\nu\theta\gamma}_{\pm}=\tfrac{1}{2}(1\pm\sqrt{1-\sin^2\alpha\sin^2\theta\sin^2\gamma/(2p^\alpha_\nu)^2})\,.
 \end{equation}
 for $\nu=\pm$. The ensuing mutual information becomes 
 \begin{equation}    \!\!\!\!\!I^{\prime\alpha}(A,T)=-\!\sum_{\nu=\pm}\lambda^\theta_\nu\log \lambda^\theta_\nu+\!\!\sum_{\nu,\mu=\pm}
     p^\alpha_\nu \lambda^{\alpha_\nu\theta\gamma}_{\mu}\log\lambda^{\alpha_\nu\theta\gamma}_{\mu}\,.\label{eqg}
 \end{equation}
 As seen in the top and bottom panels of Fig.\ \ref{fig1}, this quantity again decreases for increasing $\alpha$ in the interval $[0,\pi/2]$, but now does not vanish at $\alpha=\pi/2$ if $\cos\gamma\neq 0$, where its value increases for decreasing $\gamma$, 
in agreement with previous discussion.
In fact, for $\gamma=0$ the second sum in  \eqref{eqg} vanishes and $I^{\prime\alpha}(A,T)=S(\rho_A)=I'(A,T)$ becomes independent of $\alpha$, as  $\rho_{AT}$ becomes a pure history state $\rho_{AT}=\ketbra{\Psi_{AT}}
$, with $\ket{\Psi_{AT}}=\frac{1}{\sqrt{2}}(\ket{00}+\ket{\theta 1})$.  It is also verified in the top panel  that $I^{\prime\alpha}(A,T)-I^{\prime\alpha}(B,T)$ is independent of the measurement angle $\alpha$, in agreement with Eq.\ \eqref{dif2}.  

For $\theta=\pi/2$ ($A$ evolving between orthogonal states in the standard clock basis), $p^\alpha_{\pm}=1/2$ and Eq.\ \eqref{eqg} adopts the simpler expression 
 \begin{equation}
    I^{\prime\alpha}(A,T)=1+\sum_{\nu=\pm}     \lambda^{\alpha\gamma}_{\nu}\log\lambda^{\alpha\gamma}_{\nu}\,,\;\;\;(\theta=\pi/2)\label{eqg2}
 \end{equation}
with $\lambda^{\alpha\gamma}=\frac{1}{2}(1\pm\sqrt{1-\sin^2\alpha\sin^2\gamma})$. At the minimum $\alpha=\pi/2$, $\lambda^{\alpha\gamma}=\frac{1\pm\cos\gamma}{2}$ and hence $I^{\prime\alpha}(A,T)=1-S(\rho_B)$.

\subsubsection{Evolution of entangled states}
As a second  example, let us consider in \eqref{psiq2} 
 entangled  states $\ket{\psi^t_{AB}}$, non orthogonal in general, 
having the same local Schmidt bases  such that
 \footnote{a common phase $e^{-i\phi_t}$ can be absorbed through a replacement $\ket{t}\rightarrow e^{-i\phi_t}\ket {t}$.}
\begin{equation}
    \ket{\Psi}=\tfrac{1}{\sqrt{2}}\sum_{t=0,1}(\cos\tfrac{\theta_t}{2}\ket{00}+\sin\tfrac{\theta_t}{2}\ket{11})\ket{t}\,.\label{77}
\end{equation}
This state is symmetric so that $I^{\prime\alpha}(A,T)=I^{\prime\alpha}(B,T)$. The total system-time entanglement \eqref{EF} is determined by the overlap $\langle\psi_{AB}^0|\psi_{AB}^1\rangle=\cos\frac{\theta_0-\theta_1}{2}$. In the standard clock basis, subsystems  evolve between mixed states 
\begin{equation}
\rho_A^t=\tfrac{1}{2}(\cos^2\tfrac{\theta_t}{2}\ketbra{0}+\sin^2\tfrac{\theta_t}{2}\ketbra{1})\,,\label{78}
\end{equation} 
such that  just its eigenvalues $\frac{1\pm\cos\theta_t}{2}$  are evolving.  Hence $\rho_A=\tfrac{1}{2}\sum_t\rho_A^t$ is also diagonal in the same basis, and after a measurement in the standard clock basis, we obtain
\begin{equation}
    I'(A,T)=S(\rho_A)-\tfrac{1}{2}\sum_t S(\rho_A^t)\,,
\end{equation}
where $\rho_A=\frac{\rho_A^0+\rho_A^1}{2}$.

Changing to the general $\alpha$ basis \eqref{sal}, the ensuing conditional states \eqref{psial} have the same previous Schmidt basis, leading to the reduced conditional states 
\begin{equation}    \rho_A^{\alpha_\pm}=\cos^2\tfrac{\theta^\alpha_\pm}{2}\ketbra{0}+\sin^2\tfrac{\theta^\alpha_\pm}{2}\ketbra{1}\,,  \label{rhoac2}
\end{equation}
in \eqref{rhopatal}, 
with probabilities $p^\alpha_{\pm}=\frac{1}{2}(1\pm\cos\frac{\theta_1-\theta_0}{2}\sin\alpha)$ and 
\begin{equation}
  \begin{split}  
\cos\tfrac{\theta^\alpha_+}{2}&=(\cos\tfrac{\alpha}{2}\cos\tfrac{\theta_0}{2}+\sin\tfrac{\alpha}{2}\cos\tfrac{\theta_1}{2})/\sqrt{2p_+^\alpha}\\ \cos\tfrac{\theta^\alpha_-}{2}&=(\sin\tfrac{\alpha}{2}\cos\tfrac{\theta_0}{2}-\cos\tfrac{\alpha}{2}\cos\tfrac{\theta_1}{2})/\sqrt{2p_-^\alpha}
\end{split}\,.
\end{equation}
Then Eq.\ \eqref{ipj2} becomes 
\begin{equation}
    I^{\prime\alpha}(A,T)=S(\rho_A)+
    \sum_{\nu,\mu=\pm}    p^\alpha_{\nu}\lambda^{\alpha_\nu}_\mu\log\lambda^{\alpha_\nu}_\mu\,,\label{Ial2}
\end{equation}
where $\lambda^{\alpha_\nu}_{\pm}=\frac{1\pm\cos\theta^\alpha_\nu}{2}$. 

\begin{figure}
\centering
\includegraphics[scale=.9]{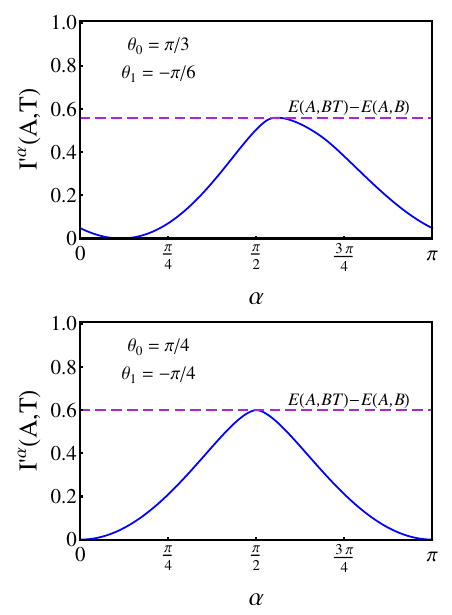}
\vspace*{-0.35cm}
\caption{Mutual information $I'^{\alpha}(A,T)$ \eqref{Ial2} (solid lines) 
 quantifying the  subsystem evolution  with respect to the clock basis $B_\alpha$, Eq.\ \eqref{sal}, for the history state \eqref{77}  at two distinct values of $(\theta_0,\theta_1)$. Maximum evolution is attained 
  for $\alpha$ given by Eq.\ \eqref{almax}, where  $I^{\prime\alpha}(A,T)$ reaches its upper limit \eqref{lem3} (horizontal dashed line), while  the minimum $0$ is attained at $\alpha$ given by Eq.\ \eqref{almin}.  
\vspace*{-0.5cm}}
\label{fig2}
\end{figure}

It is to be  noticed that 
\begin{equation}
\begin{split}    \rho_{AB}=&\cos^2\tfrac{\bar\theta}{2}\ketbra{00}+\sin^2\tfrac{\bar\theta}{2}\ketbra{11}+\\&\tfrac{\sin\theta_0+\sin\theta_1}{4}(\ketbra{00}{11}+\ketbra{11}{00})]\,,
    \end{split}
\end{equation}
with $\cos^2\tfrac{\bar\theta}{2}=\tfrac{1}{2}(\cos^2\tfrac{\theta_0}{2}+\cos^2\tfrac{\theta_1}{2})$,  is an {\it entangled} mixed state if $\sin\theta_0+\sin\theta_1\neq 0$, with concurrence 
\begin{equation}
C_{AB}=\tfrac{1}{2}|\sin\theta_0+\sin\theta_1|\,,
\end{equation}
and eigenvalues $\frac{1}{2}(1\pm\cos\tfrac{\theta_1-\theta_0}{2})$. 
This leads to an entanglement of formation $E(A,B)=\sum_{\nu=\pm}\lambda^C_\nu\log\lambda^C_\nu>0$ if $C_{AB}>0$,  with $\lambda^C_{\pm}=\frac{1}{2}(1\pm\sqrt{1-C^2_{AB}})$, and  implies that in this case the standard clock basis will not provide in general the maximum mutual information \eqref{lem3}. Instead, from Eq.\ \eqref{Ial2} it follows that the maximum is reached at 
\begin{equation}
    \tan\alpha=\frac{2\cos
\frac{\theta_0+\theta_1}{2}}{
\cos{\theta_0}-\cos{\theta_1}
}\,,
\label{almax}\end{equation}
for which the conditional  densities \eqref{rhoac2} become isospectral but  distinct ($\cos^2\tfrac{\theta^\alpha_\mp}{2}=\sin^2\tfrac{\theta^\alpha_\pm}{2}$) and \eqref{Ial2} saturates the upper limit \eqref{lem3}, 
as seen in Fig.\ \ref{fig2}. 

It is also important to note that here
\begin{equation}    \rho_{AT}=\sqrt{p_{0'}}\ketbra{00'}+\sqrt{p_{1'}}\ketbra{11'}\,,
\end{equation}
 is {\it separable}, with $\ket{0'}\propto \cos\tfrac{\theta_0}{2}\ket{0}+\cos\tfrac{\theta_1}{2}\ket{1}$,
 $\ket{1'}\propto\sin\tfrac{\theta_0}{2}\ket{0}+\sin\tfrac{\theta_1}{2}\ket{1}$  normalized clock states and $p_{0'}=(\cos^2\tfrac{\theta_0}{2}+
 \cos^2\tfrac{\theta_1}{2})/2$, $p_{1'}=1-p_{0'}$, 
 so that a strong $\alpha$-dependence of $I^{\prime\alpha}(A,T)$ is to be expected. In fact, for 
\begin{equation}
    \tan\alpha=\frac{2\sin
\frac{\theta_0+\theta_1}{2}}{
\sin{\theta_0}-\sin{\theta_1}
}\,,
\label{almin}\end{equation}
the conditional  
reduced states \eqref{rhoac2} become {\it coincident}, thus leading to $I^{\prime\alpha}(A,T)=0$.  

The behavior  of $I^{\prime\alpha}(A,T)$ with $\alpha$ is shown in Fig.\ \ref{fig2}  for 
$\theta_0=\pi/3$, $\theta_1=\theta_0/2$ (top panel) and $\theta_0=-\theta_1=\pi/4$ (bottom panel). The strong dependence with the measurement angle $\alpha$ is verified. In the first case, maximum subsystem evolution takes place for clock states \eqref{sal} with $\alpha\approx 0.56\pi$, while no subsystem evolution occurs for $\alpha\approx 0.12\pi$, whereas in the second case the maximum is reached at $\alpha=\pi/2$, while the minimum $0$ occurs at the standard clock basis ($\alpha=0$), 
as determined by Eqs.\ \eqref{almax}--\eqref{almin}, and as also obvious from Eq.\ \eqref{78} when $\theta_1=-\theta_0$ ($\rho_A^0=\rho_A^1$).   Nonetheless, in both cases the full system $AB$ is still evolving.

\subsection{Quantum Walk}
 We now apply the present formalism to  $1D$ discrete quantum walks  (QW)  \cite{Ya.93,venegasrev.12}. 
 A QW describes a quantum particle of spin $s$ moving along a $1D$ lattice, such that the system  has two degrees of freedom, position (``walker'') and spin (``coin''). Its state space is of the form $\mathcal{H}_S=\mathcal{H}_p\otimes \mathcal{H}_s$, with the (discrete) position Hilbert space $\mathcal{H}_p$ spanned by the basis $\{\ket{i}, i\in \mathbb{Z} \}$  and  $\mathcal{H}_s$ the spin-$s$ space of dimension $2s+1$. 
  We use  the present formalism 
   to quantify the evolution of the spin and position  reduced states $\rho_p(t)$ and $\rho_s(t)$ along the walk. 
   \subsubsection{Spin $\frac{1}{2}$ Quantum Walk}
 We consider first the spin $\frac{1}{2}$ QW.   
The evolution is determined by a unitary  operator
\begin{equation}
 U_{ps}=T(\mathbb{1}\otimes C)\,,   \end{equation}
 where $C$ is the ``coin'' operator,  a real traceless Hadamard-type unitary acting on the spin subspace,
\begin{equation}
C = \cos\theta\,\sigma_z+\sin\theta\,\sigma_x=
\begin{pmatrix}
  \cos \theta & \,\,\,\,\,\sin\theta \\
  \sin\theta  & - \cos \theta
   \end{pmatrix}\,,\label{coin}
\end{equation}
with the usual Hadamard coin  recovered for $\theta=\pi/4$ and the matrix indicating its representation in the standard basis $\{\ket{0}\equiv\ket{\uparrow},\ket{1}\equiv\ket{\downarrow}\}$ of $\sigma_z$ eigenstates,  whereas 

 \begin{equation}
        T=\sum_i\ketbra{i+1}{i}\otimes \ketbra{0} + \ketbra{i-1}{i}\otimes\ketbra{1}\,,
\end{equation}
 is the spin-controlled translation operator moving the walker one step forward (backward) if spin is up (down), 
such that $T(\ket{i}\ket{0})=\ket{i+1}\ket{0}$,  $T(\ket{i}\ket{1})=\ket{i-1}\ket{1}$. 

For a QW  evolving  from an initial state $|\psi_{ps}(0)\rangle=|\psi_{ps}^0\rangle\in{\cal H}_{S}$  at $t=0$, 
and associating a  state $|n\rangle\in{\cal H}_T$ of a quantum clock $T$  to each step, with $\langle n|n'\rangle=\delta_{nn'}$ and   $n=0,\ldots, N\!-\!1$, 
the QW history state for $N$ steps is  \cite{LBCR.22} 
\begin{equation} |\Psi\rangle =\frac{1}{\sqrt{N}}\sum_{n=0}^{N-1}|\psi_{ps}^n\rangle|n\rangle,
\label{HS05}\end{equation}
where 
\begin{equation}|\psi_{ps}^n\rangle=U^n_{ps}|\psi_{ps}^0\rangle\label{Un}\end{equation} 
is the state after $n$ steps. 

We consider product initial states  $|\psi_{ps}^0\rangle=|\psi_p^0\rangle|\chi_s^0\rangle$ with $|\psi_p^0\rangle=|i_0\rangle$ a localized position state (walker at site $i_0$) 
and $|\chi_s^0\rangle=\cos\gamma|0\rangle+\sin\gamma |1\rangle$ a pure initial spin such that $\langle\bm\sigma\rangle_0=(\sin 2\gamma, 0, \cos 2\gamma)$.  
Here we focus on the evolution of the reduced states $\rho^n_{p(s)}={\rm Tr}_{s(p)}\,\ketbra{\psi_{ps}^n}$, which become mixed  for $n\geq 1$. 

The exact analytic solution for $|\psi^n_{ps}\rangle$ can be obtained through a discrete Fourier transform in position \cite{LBCR.22}. For $n\leq M$ steps it leads to  $U_{ps}^n=\sum_{k=0}^{M-1} \ketbra{k}\otimes U_k^n$,  
with $|k\rangle=\frac{1}{\sqrt{M}}\sum_{i=0}^M e^{\imath 2\pi ik/M}|i\rangle$,  
$U_k^n=e^{-in\omega_k}\ketbra{s_k^+}-e^{in\omega_k}\ketbra{s_k^-}$ and 
$|s_k^\pm\rangle$ the eigenstates of $e^{-i2\pi k\sigma_z/M}C$, with $\pm e^{\mp i\omega_k}$ its eigenvalues \cite{LBCR.22}. 
Hence the  Kraus operators  in \eqref{Kr}-\eqref{Eks} are  $E^s_k=\langle k|i_0\rangle U_k^n$ for the spin subsystem and 
$E_j^p=\sum_k|k\rangle\langle k|\langle j|U^n_k|\chi_s^0\rangle$, $j=0,1$, 
for position. 

\begin{figure}[ht]
\centering
\includegraphics[scale=0.8]{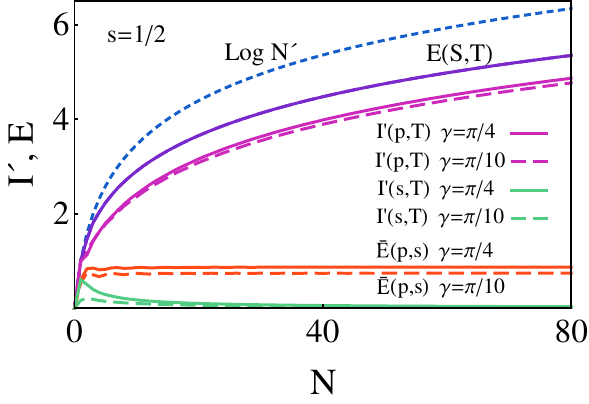}
\vspace*{-0.35cm}
\caption{Measures of total ($E$) and subsystem ($I'$) evolution as a function of the number of steps $N$ in the spin $1/2$ QW  for an initially localized particle with two  different initial spin orientations  $\gamma=\pi/10$ (solid lines) and $\gamma=\pi/4$ (dashed lines).  The evolution of the whole system is quantified  by the system-time entanglement entropy  $E(S,T)$, independent of $\gamma$, whereas that  of the position and spin reduced states  by the mutual informations   $I'(p,T)$ and $I'(s,T)$ respectively (Eq.\ \eqref{Hol}).  The average position-spin entanglement $\bar{E}(p,s)$ in the first $N$ steps (second term in \eqref{IpiAT0}-\eqref{IpiAT00}) is also shown. 
The upper bound ${\log N'}$ ($N'=N+1$) for $E(S,T)$ is as well indicated.
We have set $\log=\log_2$ in all quantities.}
\label{fig3}
\end{figure}

In  Fig.\ \ref{fig3} we show, for the Hadamard coin, the evolution of the whole system-time entanglement entropy $E(S,T)$ and the  subsystem-time mutual informations $I'(p,T)$, $I'(s,T)$ for the position and spin subsystems, 
 quantifying the evolution of $\ket{\psi^n_{ps}}$, $\rho^n_{p}$ and $\rho^n_s$ respectively,  for two different initial spin orientations  $\gamma=\pi/10$  and $\gamma=\pi/4$. 
As demonstrated in \cite{LBCR.22}, 
 $E(S,T)$ is independent of the initial spin orientation for the coin \eqref{coin} and any real initial state (with definite site parity).

It is first seen that $I'(p,T)$ follows the behavior of $E(S,T)$,  with  $I'(p,T)\approx 0.9 E(S,T)$ at $N\approx 80$, verifying the upper bound \eqref{desg}, although it exhibits a weak dependence on the initial spin orientation angle $\gamma$, absent in $E(S,T)$. Besides, $E(S,T)$ lies below the maximum value $E_{\rm max}=\log (N+1)$ reached for maximum evolution onto a new orthogonal state at each step ($E(S,T)\approx 0.85 E_{\rm max}$ at $N=80$), indicating a non vanishing overlap  between the evolved states $|\psi_{ps}^n\rangle$ and hence $\rho_p^n$. Nonetheless, the approximate proportionality with $E_{\rm max}$ reflects  a ``steady'' evolution involving new orthogonal components at each step.

In contrast, $I'(s,T)$, which cannot exceed $\log 2$ ($1$ in present units) and also exhibits a weak dependence on $\gamma$, decreases as $N$ increases after the first initial steps, since the spin reduced states $\rho^n_s$  approach for these $\gamma$ values an oscillatory behavior  between closely lying reduced states. We also depict the average 
position-spin entanglement entropy up to step $N$, which approaches the average entropy of previous asymptotic reduced states, and affects both $I'(p,T)$ and $I'(s,T)$ (second term in Eq.\ \eqref{IpiAT00}).   

\begin{figure}[ht]
\centering
\includegraphics[scale=0.7]{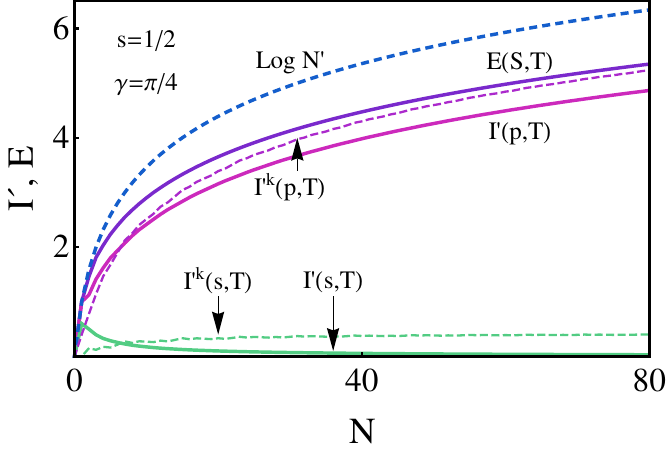}
\vspace*{-0.35cm}
\caption{The subsystem-time mutual informations $I'^k(p,T)$ and $I'^k(s,T)$ (dashed lines) quantifying the evolution with respect to the Fourier clock basis, as  a function of the number of steps $N$ in the system of Fig.\ \ref{fig3} for $\gamma=\pi/4$. 
Results for the standard clock basis $I'(p,T)$, $I'(s,T)$ (solid lines) are also depicted for comparison, together with $E(S,T)$, which is independent of the clock basis.}\label{fig4}
\end{figure}

In Fig.\ \ref{fig4} we show illustrative results for the subsystem evolution with respect to a different clock basis for the same global history state, according to the discussion of sec.\  \ref{II.E}. 
We compare previous mutual informations  $I'(p,T)$ and $I'(s,T)$  in the standard clock basis $\{|n\rangle\}$ with those obtained in the corresponding Fourier clock basis $\{|k\rangle=\frac{1}{\sqrt{N}}\sum_{n=0}^{N-1}e^{-i2\pi nk/N}|n\rangle\}$, $I'^k(s,T)$ and $I'^k(p,T)$.   
 It is seen that there is an appreciable difference between the evolutions of each subsystem for different clock bases, and that the largest one  corresponds to the Fourier clock basis, where  $I'^k(p,T) \approx 0.98 E(S,T)$ at $N\approx 80$. 
 
 In particular, the spin mutual information $I'^k(s,T)$  now approaches a constant value as $N$ increases, reflecting a roughly oscillatory behavior between non orthogonal yet different reduced states.  The whole system-time entanglement entropy $E(S,T)$  remains of course unchanged and provides an upper bound to the subsystem-time mutual information  with respect to any clock basis (Eq.\ \eqref{desg}).

\subsubsection{Spin $1$ Quantum Walk}

\begin{figure}[h] 
\centering
\includegraphics[scale=1.]{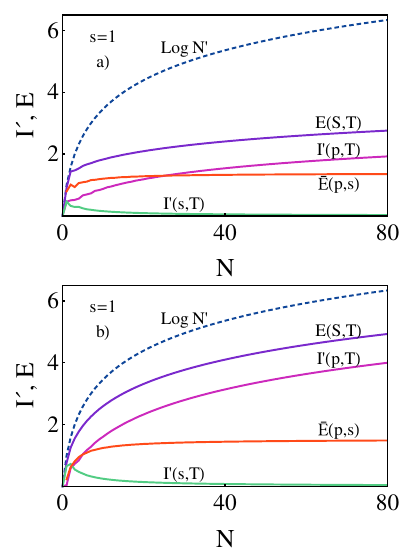}
\vspace*{-0.35cm} 
\caption{The same quantities as in Fig.\ \ref{fig3} in the spin $1$ QW, quantifying the whole ($E(S,T)$) and subsystems ($I'(p,T)$ for position,  $I'(s,T)$ for spin)  evolutions, for an initially localized particle with two different initial spin states (see text). The time average of the position-spin entanglement, $\bar{E}(p,s)$,  is also depicted.}
\label{fig5}
\end{figure}

We consider now the spin $1$ case, in which the spin state space is spanned by the three eigenstates $\{\ket{1},\ket{0},\ket{-1} \}\ $, where  $\ket{1}\equiv\ket{\uparrow}$ 
($\ket{-1}\equiv\ket{\downarrow}$) leads to a forward  (backward) step while the new eigenstate $\ket{0}$ allows the walker to stay in place: 
\begin{equation}
\begin{split} T=&\sum_i(\ketbra{i\!+\!1}{i}\otimes \ketbra{1} + \ketbra{i\!-\!1}{i}\otimes\ketbra{-1}\\
    &+\ketbra{i}\otimes \ketbra{0})\,.
    \end{split}
\end{equation}
We employ the natural generalization of the Hadamard coin, denoted as Grover coin \cite{Konno.05,Konno2.05},
\begin{equation}
C_G = \tfrac{1}{3}
\begin{pmatrix}
  -1 & \,\,\, 2 &\,\,\,  2  \\
 \,\,\, 2 & -1 & \,\,\, 2  \\
  \,\,\, 2 & \,\,\, 2 & -1
 \end{pmatrix}
\end{equation}
satisfying $C_G^\dag C_G=\mathbbm{1}$, which yields identical probabilities of moving left, right, or staying, when applied to a uniform equal phase spin state 
$|\chi^0_s\rangle=\frac{|1\rangle+|0\rangle +|-1\rangle}{\sqrt{3}}$. 

The evolution of the whole system is determined by the operator $U=T(\mathbbm 1 \otimes C_G)$. The QW history state is given again by Eqs.\  \eqref{HS05}--\eqref{Un}, with  $|\psi_{ps}^0\rangle=|\psi_p^0\rangle\otimes |\chi_s^0\rangle$, where $|\psi_p^0\rangle=|i_0\rangle$ 
and  $|\chi^0_s\rangle=\alpha|-1\rangle+\beta |0\rangle+\gamma |1\rangle$.

In  Fig.\ \ref{fig5} we show the system and subsystem evolutions  quantified by  $E(S,T)$, $I'(p,T)$ and  $I'(s,T)$ respectively, as a function of the number of  steps $N$  for two different initial spin conditions: a) $\ket{\chi^0_s}=\frac{\ket{1}+\ket{0}+\ket{-1}}{\sqrt{3}}$ (top panel), and  b)  $\ket{\chi^0_s}=\frac{\ket{1}-\ket{0}+\ket{-1}}{\sqrt{3}}$ (bottom panel).  
 We notice that in the spin $1$ QW  the evolution of all quantifiers, including $E(S,T)$, depend on the initial conditions. The chosen initial conditions are representative of those leading to minimum and maximum evolution for uniform weight coefficients.
 
 It is again seen that $I'(p,T)$ follows the behavior of $E(S,T)$, with $I'(p,T)\approx 0.7 E(S,T)$ in case a) and $I'(p,T)\approx 0.8 E(S,T)$ in case b) at $N\approx 80$, 
 fulfilling Eq.\ \eqref{desg}. Nonetheless, $E(S,T)$ is now smaller than before due to the possibility of staying at the same location: $E(S,T)\approx 0.43 E_{\rm max}$ in case a) and $E(S,T)\approx 0.78 E_{\rm max}$ in b). 
On the other hand, $I'(s,T)$ again decreases as $N$ increases since 
the spin reduced state evolution exhibits a small amplitude oscillatory behavior between very close states.  The average 
position-spin entanglement entropy is also shown,  and exhibits a slight dependence  on the initial conditions,  as in  the spin $1/2$ case shown in Fig. \ref{fig3}, approaching the average entropy  of the asymptotic spin states. This quantity affects both $I'(p,T)$ and $I'(s,T)$ through  Eqs.\ \eqref{IpiAT0}-\eqref{IpiAT00}).

\section{Conclusions}\label{sec:conclusions}

We have introduced, in the context of the history state formalism, a general quantifier of distinguishable subsystem evolution, based on 
a subsystem-clock mutual information. 
For a measurement in the standard clock basis, $I'(A,T)$, Eq.\ \eqref{Hol},  quantifies the  evolution of the reduced subsystem state $\rho_A(t)$ in the history, representing the logarithm of the  effective number of orthogonal reduced states  visited. While applicable to any set of mixed states $\rho_A(t)$, in the context of the composite history state it acquires a clear  meaning as the difference between the total $A$--$BT$ entanglement  and the average $A$--$B$ entanglement along the history,  being bounded above by the total system-time entanglement entropy $E(S,T)$. It holds for arbitrary initial states $|\psi_{AB}(0)\rangle$, such that when entangled, 
the evolution of $\rho_A(t)$ is governed  in general by a full  matrix of Kraus operators ${\bf E}(t)$ through Eq.\ \eqref{EMF},   the standard CPTP recovered for product initial states. 

Furthermore,  the history state scenario admits the consideration of different clock bases for a given history, and reveals that  the degree of subsystem evolution actually depends on the choice of the reference clock basis, in contrast with that of the full system. 
This naturally leads to an ``optimal'' clock basis which maximizes the subsystem evolution quantifier $\tilde I'(A,T)$, whose   maximum is just the difference    \eqref{lem3} between the total $A$--$BT$ entanglement and the $A$--$B$ 
 entanglement of formation 
 in the history state. On the other hand,  the subsystem-clock quantum entanglement of formation $E(A,T)$ in the history,  
while not directly a measure of the subsystem reduced state evolution, 
 affects the  sensitivity of the  clock basis dependence of $\tilde I'(A,T)$: this basis dependence vanishes for a pure entangled $\rho_{AT}$ and peaks when  $\rho_{AT}$ is classically correlated from $T$ (hence $E(A,T)=0$), in which case there is a clock basis where the subsystem reduce state appears stationary  ($\tilde I'(A,T)=0$). We illustrated these features analytically  in  
   simple qubit-clock examples, where it was explicitly  demonstrated that null subsystem evolution with respect to some clock basis implies $E(A,T)=0$. 

Finally, we demonstrated the practical utility of this framework by consistently quantifying both the whole system evolution as well as the spin and position reduced state evolutions in $1D$ spin $1/2$ and spin $1$ quantum walks, within a unified formalism. Extending these concepts to characterize broader classes of open quantum system dynamics remains a promising avenue for future investigation.

\acknowledgments
Authors acknowledge support from CONICET (F.L. and N.C.), Departamento de Física UNLP (A.B.) and CIC (R.R.) of Argentina.  Work supported by
CONICET PIP Grant No. 11220200101877CO.

\appendix
\section{History-state representation}
We show that {\it any} bipartite state $\ket{\Psi}\in{\cal H}_S\otimes {\cal H}_T$  can be written as a history state \eqref{HS},   with the states $\ket{\psi_S(t)}$ undergoing a unitary evolution $\ket{\psi_S(t)}=U(t)\ket{\psi_S(0)}$. We start  from its Schmidt decomposition, 
\begin{equation}    |\Psi\rangle=\sum_{k=0}^{n_s-1}\sqrt{p_k}|k_S\rangle |k_T\rangle\label{SDA}
\end{equation}
where $|k_{S(T)}\rangle$ are orthogonal states of subsystems $S$ and $T$: $\langle k_{S(T)}|k'_{S(T)}\rangle=\delta_{kk'}$, and  $p_k>0$, $\sum_{k=0}^{n_s-1} p_k=1$, with  $n_s\leq {\rm Min}[d_S,d_T]$  its Schmidt rank. We may consider the $n_s$ states $|k_T\rangle$ as forming part of an orthogonal set of $N\leq d_T$  states $|k_T\rangle$, with $N\geq n_s$. Then we  define their discrete Fourier transformed states $|t\rangle=\frac{1}{\sqrt{N}}\sum_{k=0}^{N-1}e^{i2\pi kt/N}|k_T\rangle$ for $t=0,\ldots,N-1$, satisfying 
$\langle t|t'\rangle=\delta_{tt'}$, 
such that \begin{equation}|k_T\rangle=\frac{1}{\sqrt{N}}\sum_{t=0}^{N-1}e^{-i2\pi kt/N}|t\rangle\,.\label{FT}\end{equation} 
The state \eqref{SDA} can then be recast as a  history state 
\begin{equation}
    |\Psi\rangle=\frac{1}{\sqrt{N}}\sum_{t=0}^{N-1} |\psi_{S}(t)\rangle|t\rangle\,,\label{HS}
\end{equation}
where 
\begin{equation}
    |\psi_{S}(t)\rangle=\sum_{k=0}^{n_s-1} \sqrt{p_k}|k_S\rangle e^{-i2\pi kt/N}=U(t)\ket{\psi_S(0)}\,,
    \label{psist}
\end{equation}
are normalized states, $\langle \psi_S(t)|\psi_S(t)\rangle=\sum_{k=0}^{n_s-1}p_k=1$, and can be seen as  evolving from $|\psi_S(0)\rangle=\sum_{k=0}^{n_s-1}\sqrt{p_k}|k_S\rangle$ through a unitary operator $U(t)=e^{-iH_S t}$,  with  $H_S$ having the states $\ket{k_S}$ as eigenstates: $H_S|k_S\rangle=\frac{2\pi}{N}|k_S\rangle$. 

Previous scheme can be extended to more general  unitary operators: Defining e.g.\  $\ket{ t}=\sum_{k=0}^{N-1} W_{kt}^*\ket{k_T}$, 
with   $W_{kt}=e^{-i\phi_{kt}}/\sqrt{N}$  the elements of a complex $N\times N$ unitary Hadamard matrix $W$ (hence  
 satisfying $|W_{kt}|=1/\sqrt{N}\,\forall\,k,t$), we still obtain 
$\langle  t| t'\rangle=\delta_{tt'}$  and  
$\ket{k_T}=\sum_{t=0}^{N-1}W_{kt}\ket{t}$. We then arrive again at a  representation \eqref{HS} of $\ket{\Psi}$, where 
 \begin{equation}|\psi_S(t)\rangle=\sum_k e^{-i\phi_{kt}}\sqrt{p_k}|k_S\rangle\,\end{equation} are still normalized  $\forall\,t$: $\braket{\psi_S(t)}=\sum_k  p_k=1$,  hence connected to  $|\psi_S(0)\rangle$ through a  unitary transformation $\ket{\psi_S(t)}=U(t)\ket{\psi_S(0)}$, with 
$U(t)$ again having the states $\ket{k_S}$ as eigenstates: $U(t)\ket{k_S}=e^{-i(\phi_{kt}-\phi_{k0})}\ket{k_S}$.

\section{No subsystem-clock correlation and separability}
Here we prove the statement   of Section III.1: In a  general two-qubit two-times scenario,  if  $I^{\prime\alpha}(A,T)=0$, i.e. if there is no subsystem evolution with respect to  some orthogonal clock basis  $B_\alpha$,  the original reduced state $\rho_{AT}$ {\it is separable}, i.e., it has 0 entanglement of formation, though it is not necessarily classically correlated.  

{\it Proof:} 
Changing for convenience of notation the order of $A,T$, 
we prove that if a mixed state  $\rho_{_{TA}}$ yields identical local states $\rho_A$ after measurement in a given clock basis $B_\alpha=\{|\alpha_\pm\rangle\}$ with $\langle \alpha_+|\alpha_-\rangle=0$, 
such that  $\langle \alpha_\pm|\rho_{T,A}|\alpha_\pm\rangle=p_{\pm}\rho_A$, with $p_++p_-=1$,  then $\rho_{_{TA}}$ has positive partial transpose and is then  separable. 

The previous condition implies that in a product basis $B_\alpha\otimes B_A$, $\rho_{_{TA}}$ adopts the form 
\begin{equation}\rho_{_{TA}}=
\begin{pmatrix}p_{_+}\rho_A&\sigma\\\sigma^\dag&p_{_-}\rho_A\end{pmatrix}
\end{equation}
where $\sigma$ is a  $2\times 2$ matrix and $\rho_A$ can be considered real and diagonal  by choosing $B_A$ as that formed by its eigenvectors. Its partial transpose is just 
\begin{equation}\rho_{_{TA}}^{t_A}=\begin{pmatrix}p_{_+}\rho_A&\sigma^t\\\sigma^*&p_{_-}\rho_A\end{pmatrix}
\end{equation}
If $p_+$ or $p_-$ vanishes or if $\rho_A$ has rank $1$, $\rho_{_{TA}}$ is trivially separable, since its positivity 
implies in this case $\sigma=0$ or $\sigma$ of rank $1$ respectively (if e.g.\  $\rho_A=(^{0\;0}_{0\;1})$ then $\sigma\propto \rho_A$) and  hence $\rho^{t_A}_{_{TA}}=\rho_{_{TA}}$. 
On the other hand, if $p_{\pm}\rho_A$ are positive definite,   we define
\begin{subequations}
\begin{eqnarray}
\tilde\rho_{_{TA}}&=&D^{-1/2}\rho_{_{TA}}D^{-1/2}=  
\begin{pmatrix}\mathbbm 1&\tilde \sigma\\\tilde \sigma^\dag&\mathbbm{1}\end{pmatrix}\\
\tilde\rho_{_{TA}}^{t_A}&=&
D^{-1/2}\rho_{_{TA}}^{t_A}D^{-1/2}=  
\begin{pmatrix}\mathbbm 1&\tilde \sigma^t\\\tilde \sigma^*&\mathbbm{1}\end{pmatrix}
\end{eqnarray}
\label{rtd}
\end{subequations}
where $\tilde\sigma=\frac{1}{\sqrt{p_+p_-}}\rho_A^{-1/2}\sigma \rho_A^{-1/2}$ and
\begin{equation}
 D= 
\begin{pmatrix}p_+\rho_A&0\\0&p_-\rho_A\end{pmatrix}\,. 
\end{equation}
The matrices \eqref{rtd} are isospectral since they are hermitian and related by $\tilde\rho_{_{TA}}^{t_A}=P\tilde\rho_{_{TA}}^*P$ with $P=(^{0\;\mathbbm 1}_{\mathbbm 1\;0})=\sigma_x\otimes \mathbbm{1}=P^{-1}$.  
As $D$ is positive definite, positivity of $\rho_{_{TA}}$ implies that of $\tilde\rho_{_{TA}}$ and hence that of $\tilde\rho_{_{TA}}^{t_A}$ and $\rho^{t_A}_{_{TA}}$, entailing separability of $\rho_{_{TA}}$ by the Peres-Horodecki criterion \cite{P.96, Ho.96}. 
Thus $E(A,T)>0$ (i.e., $C_{AT}>0$) prevents $I^{\prime\alpha}(A,T)$ from vanishing in any orthogonal clock basis. 

%\bibliography{biblio}


\end{document}